\documentstyle[righttag,amscd,amssymb,verbatim,12pt]{amsart}
\theoremstyle{plain}
\newtheorem{Thm}{Theorem}[section]
\newtheorem{Cor}{Corollary}[section]
\newtheorem{Lem}{Lemma}[section]
\newtheorem{Prop}{Proposition}[section]
\newtheorem{MainLem}{Main Lemma}

\newtheorem{ThmB}{Theorem {[B]} }

\newtheorem{PropH}{Proposition {[H, page 50]} }

\theoremstyle{definition}

\theoremstyle{remark}
\newtheorem{Rem}{Remark}

\renewcommand{\rm}{\normalshape}
\numberwithin{equation}{section}

\def\bx{\bold x} 
\def\bz{\bold Z}
\def\av{\alpha^{\vee}}   
\def\sg{\sigma}
\def\al{\alpha}
\def\w{\omega}
\def\baf{\bar f}
\def\bc{\bold C}
\def\tw{\widetilde W}
\def\tv{\widetilde V}
\def\ty{\tilde y}
\def\zy{y^{0}}  
\def\ts{\text{T}^*\cal M}
\def\bn{\nabla}

\def\ve{\varepsilon}

\def\pal{\partial}   
\def\lm{\lambda}
\def\res{\mathop{\text{\rm res}}}
\def\ta{\theta} 
\begin{document}

\title[Affine Weyl Groups and Frobenius manifolds]
{Extended affine Weyl groups and Frobenius manifolds}
\author[B. Dubrovin, Y.J. Zhang]{Boris Dubrovin$^{1,2}$,
       Youjin Zhang$^{1}$}
\address[mailing address]{SISSA\\ Via Beirut 2--4\\ 34014 Trieste\\ Italy}
\email{dubrovin@@sissa.it\\ \ youjin@@sissa.it}
\thanks{$^1$ SISSA, Via Beirut 2--4, 34014 Trieste, Italy}
\thanks{$^2$ Steklov Math. Institute, Moscow}
\maketitle
\begin{abstract}
We define certain extensions of affine Weyl groups 
(distinct from these considered by K. Saito [S1] in the
theory of extended affine root systems), 
prove an
analogue of Chevalley theorem
for their invariants, and construct a Frobenius structure
on their orbit spaces. This produces solutions $F(t_1, \dots, t_n)$
of WDVV equations of associativity polynomial in $t_1$, \dots, $t_{n-1}$, 
$\exp t_n$.\par
\end{abstract}

\section{Introdution}
Frobenius manifold is a geometric object (see precise definition in
Section 2 below) designed as a coordinate-free formulation of
{\it equations of associativity}, or {\it WDVV equations} (they were
invented in the beginning of '90s by Witten, Dijkgraaf, E. and H. 
Verlinde in the setting of two dimensional topological field theory; 
see [D] and references therein). In [D] for an arbitrary 
$n$-dimensional Frobenius manifold a {\it monodromy group} was 
defined. It acts in $n$-dimensional linear space and it is an 
extension of a group generated by reflections. Looking at simple 
examples it might be conjectured that for a Frobenius manifold with 
good analytic properties (in the sense of [D], Appendix A) 
the monodromy group acts discretely in some 
domain of the space. The Frobenius manifold itself can be identified 
with the orbit space of the group in the sense to be specified for 
each class of monodromy groups.\par
In the present paper we introduce a new class of discrete groups that 
can be realized as monodromy groups of Frobenius manifolds (it was 
shown previously that any finite Coxeter group can serve as a 
monodromy group of a polynomial Frobenius manifold, see [D]). We 
define certain extensions of affine Weyl groups, and construct a 
Frobenius structure on their orbit spaces. Our groups coincide with 
the monodromy groups of the Frobenius manifolds. They are labelled
by pairs $(R,k)$ where $R$ is an irreducible reduced root system, and 
$k$ is a certain simple root (shown in white on Table 1 below). 
Our construction of Frobenius structure includes, particularly,
a construction of the flat coordinates $\, t_1,\dots,t_n$\, in 
the appropriate ring of invariants of the extended affine Weyl 
groups (flat coordinates in the ring of polynomial invariants
of finite Coxeter groups were discovered by Saito, Yano, Sekiguchi 
[SYS, S]). The correspondent 
solutions of equations of associativity are weighted
homogeneous (up to a quadratic function) polynomials 
in $t_1, \dots, t_{n-1}$, $e^{t_n}$ with all positive weights of the 
variables. Here $n-1$ is equal to the rank of the root system $R$. It 
can be shown (see [D], Appendix A) that for $n\leq 3$ our 
construction exhausts all such solutions.\par
The paper is organized in the following way. In Section 1 we define 
extended affine Weyl groups and prove an analogue of Chevalley 
theorem [B] for them. In Section 2 we construct Frobenius structure 
on the orbit spaces of our groups and compute explicitly all 
low-dimensional examples of these Frobenius manifolds. In Section 3 
we show that in the case of the root system of $A$-type, our extended 
affine Weyl groups describe monodromy of roots of trigonometric 
polynomials of a given bidegree. We discuss topology of the complement
to bifurcation variety of such trigonometric polynomials in terms of 
the correspondent Frobenius manifolds.\par

\section{Extended affine Weyl groups and their invariants}
Let $R$ be an irreducible reduced root system in $l$-dimensional
Euclidean space $V$ with Euclidean inner product $(~,~)$. 
We fix a basis $\ \al_1,\al_2,\dots,\al_l$\
of simple roots. Let
$$
\av_j=\frac {2\al_j}{(\al_j,\al_j)},\quad j=1,2,\dots,l
$$
be the correspondent coroots. All the numbers $\ A_{ij}:=(\al_i,\av_j)$\
are integers  (these are the entries of the Cartan matrix $\ A=\left( 
A_{ij}\right)$, $\
(\al_i,\av_i)=2,\quad (\al_i,\av_j)\le 0\ \ \text {for} \ \
i\ne j$).\ The Weyl group $\ W=W(R)$\ is a finite group generated
by the reflections $\ \sg_1,\sg_2,\dots,\sg_l$:
\begin{equation}
\sg_j(\bx)=\bx-(\av_j,\bx)\al_j, \quad\bx\in V.\tag{1.1}
\end{equation}
We recall that the root system is one of the type $\ A_l,\ B_l,\
C_l,\ D_l,\ E_6,\ E_7,\ E_8,\ F_4,\ G_2$\ (see [B]).\par
The affine Weyl group $\ W_a(R)$\ acts in the space $\ V$\ by
affine transformations
$$
\bx\mapsto w(\bx)+\sum_{j=1}^l m_j\av_j,\quad w\in W,\ m_j\in \bold Z.
$$
So it is isomorphic to the semidirect product of $\ W$\ by the lattice
of coroots.\par
Let us introduce coordinates $\ x_1,x_2,\dots, x_l$ in $\ V$\
using the basis of coroots:
\begin{equation}
\bx=x_1\av_1+x_2\av_2+\cdots+x_l\av_l.\tag{1.2}
\end{equation}
We define {\it {Fourier polynomial}} as the following functions on
$\ V$:
$$
f(\bx)=\sum_{m_1,\dots,m_l\in\bz} a_{m_1,\dots,m_l} e^{2\pi i
(m_1 x_1+\cdots+m_l x_l)},
$$
the coefficients are arbitrary complex numbers and only finite number
of them could be nonzero. Alternatively, introducing the fundamental
weights $\ \w_1,\dots,\w_l \in V$,
$$
(\w_i,\av_j)=\delta_{ij},
$$
we can represent the Fourier polynomial as a sum over the
weight lattice:
$$
f(\bx)=\sum_{m_1,\dots,m_l\in\bz} a_{m_1,\dots,m_l} e^{2\pi i
(m_1 \w_1+\cdots+m_l\w_l,\bx)}.
$$
Thus the ring of our Fourier polynomials is identified with the group
algebra of the weight lattice [B].
We define the operation of averaging of a Fourier polynomial
\begin{equation}
f(\bx)\mapsto \baf(\bx)=S_W(f(\bx)) :=
n_f^{-1}\,\sum_{w\in W}f(w(\bx)),\tag{1.3}
\end{equation}
where $\, n_f=\# \{w\in W\,|\, f(w(\bx))=f(\bx)\}$.\
For any $\ f(\bx)$\ the Fourier polynomial
$\ \baf(\bx)=S_W(f(\bx))$\ is a function on $\ V$\ invariant with
respect to the action of the affine Weyl group:
$$
\baf(w(\bx)+\sum_{j=1}^l m_j\av_j)=\baf(\bx).
$$
Equivalently, this is a $W$-invariant Fourier polynomial.\par
\begin{ThmB} The ring of $\ W$-invariant Fourier
polynomials is isomorphic to the polynomial ring
$\ \bc[y_1,\dots,y_l]$,\ where $\ y_1=y_1(\bx),\dots,
y_l=y_l(\bx)$\ are the basic $\ W$-invariant Fourier polynomials
defined by
\begin{equation}
y_j=S_W(e^{2\pi i (\w_j,\bx)}),\quad j=1,\dots,l.\tag{1.4}
\end{equation}
\end{ThmB}
\noindent {\bf {Example 1.1.}}\ \ The Weyl group $\ W(A_l)$\ acts by
permutations of the coordinates \newline $z_1,\dots,z_{l+1}$ \
on the hyperplane
$$
z_1+\cdots+z_{l+1}=0.
$$
We choose the standard basis of simple roots $\ \al_j=\av_j$\
as in [B, Planches I]. Then the coordinates $\ x_1,\dots,
x_l$\ are defined by
\begin{equation}
z_1=x_1,\quad z_i=x_i-x_{i-1},\quad i=2,\dots,l,\quad  z_{l+1}=-x_l.
\tag{1.5}
\end{equation}
The basic $W$-invariant Fourier polynomials coincide with the
elementary symmetric functions
\begin{equation}
y_j=s_j(e^{2\pi i z_1},\dots,e^{2\pi i z_{l+1}}),\quad
j=1,\dots,l.\tag{1.6}
\end{equation}
\par
We are going now to define certain extensions of the affine
Weyl group acting in the $(l+1)$-dimensional space with
indefinite metric.\par
For any irreducible reduced root system $\ R$\ we fix 
a root $\ \al_k$\ 
indicated in 
Table 1. The Dynkin graphs of $\ A-B-C-D-E-F-G$\ \ 
type are shown in the
Table 1 with one more vertex added (this is indicated by
asterisk). We will use this additional vertex later on.
The white vertex of the Dynkin graph corresponds to the chosen root
$\ \al_k$.\ 
Observe that the Dynkin graph of 
$\ R_k :=\{\al_1,\dots,\hat {\al}_k,\dots,\al_l\}\ (\al_k$\ is 
\par
\begin{figure}
\setlength{\unitlength}{0.1in}%
\begin{picture}(60,80)(-4,-40)
\put(0,34.5){\makebox(2,1)[l]{$A_l$}}
\put(5,35){\line(1,0){5}}
\put(5,35){\circle * {0.44}}
\put(4,33.2){\makebox(2,1)[o] {\footnotesize {$1$}}}
\put(10,35){\circle * {0.44}}
\put(9,33.2){\makebox(2,1)[o] {\footnotesize {$2$}}}
\put(12,35){\circle * {0.25}}
\put(14,35){\circle * {0.25}}
\put(16,35){\circle * {0.25}}
\put(18,35){\line(1,0){4.75}}
\put(23.25,35){\line(1,0){4.8}}
\put(18,35){\circle * {0.44}}
\put(17,33.2){\makebox(2,1)[o] {\footnotesize {$k-1$}}}
\put(23,35){\circle{0.5}}
\put(22,33.2){\makebox(2,1)[o] {\footnotesize {$k$}}}
\put(28,35){\circle * {0.44}}
\put(27,33.2){\makebox(2,1)[o] {\footnotesize {$k+1$}}}
\put(30,35){\circle * {0.25}}
\put(32,35){\circle * {0.25}}
\put(34,35){\circle * {0.25}}  
\put(36,35){\line(1,0){5}}
\put(36,35){\circle * {0.44}}
\put(35,33.2){\makebox(2,1)[o] {\footnotesize {$l-1$}}}
\put(41,35){\circle * {0.44}}
\put(40,33.2){\makebox(2,1)[o] {\footnotesize {$l$}}}
\put(41,35){\line(1,0){5}}
\put(45.8,34.8){\makebox(0,0)[o] {\bf *}}
\put(45,33.2){\makebox(2,1)[o] {\footnotesize {$l+1$}}}
\put(0,29.5){\makebox(2,1)[l]{$B_l$}}
\put(5,30){\line(1,0){5}}   
\put(5,30){\circle * {0.44}}
\put(4,28.2){\makebox(2,1)[o] {\footnotesize {$1$}}}
\put(10,30){\circle * {0.44}}
\put(9,28.2){\makebox(2,1)[o] {\footnotesize {$2$}}}
\put(12,30){\circle * {0.25}}
\put(14,30){\circle * {0.25}}
\put(16,30){\circle * {0.25}}
\put(18,30){\circle * {0.44}}
\put(17,28.2){\makebox(2,1)[o] {\footnotesize {$k-1$}}}
\put(18,30){\line(1,0){5}}
\put(23,30){\line(1,0){5}}
\put(23,30){\circle * {0.44}}   
\put(22,28.2){\makebox(2,1)[o] {\footnotesize {$k$}}}
\put(28,30){\circle * {0.44}}
\put(27,28.2){\makebox(2,1)[o] {\footnotesize {$k+1$}}}  
\put(30,30){\circle * {0.25}}
\put(32,30){\circle * {0.25}}
\put(34,30){\circle * {0.25}}
\put(35.8,29.8){\line(1,0){5.1}}
\put(35.8,30.2){\line(1,0){5.1}}
\put(36,30){\circle{0.5}}
\put(35,28.2){\makebox(2,1)[o] {\footnotesize {$l-1$}}}
\put(41,30){\circle * {0.44}}
\put(37.5,29.493){\makebox(2,1)[o] {$>$}}
\put(40,28.2){\makebox(2,1)[o] {\footnotesize {$l$}}}
\put(41,30){\line(1,0){5}}
\put(45.8,29.8){\makebox(0,0)[o] {\bf *}}
\put(45,28.2){\makebox(2,1)[o] {\footnotesize {$l+1$}}}
\put(0,24.5){\makebox(2,1)[l]{$C_l$}}  
\put(5,25){\line(1,0){5}}
\put(5,25){\circle * {0.44}} 
\put(4,23.2){\makebox(2,1)[o] {\footnotesize {$1$}}}
\put(10,25){\circle * {0.44}}
\put(9,23.2){\makebox(2,1)[o] {\footnotesize {$2$}}}
\put(12,25){\circle * {0.25}}
\put(14,25){\circle * {0.25}}
\put(16,25){\circle * {0.25}}
\put(18,25){\circle * {0.44}}
\put(17,23.2){\makebox(2,1)[o] {\footnotesize {$k-1$}}}
\put(18,25){\line(1,0){5}}
\put(23,25){\line(1,0){5}}
\put(23,25){\circle * {0.44}}
\put(22,23.2){\makebox(2,1)[o] {\footnotesize {$k$}}}  
\put(28,25){\circle * {0.44}}
\put(27,23.2){\makebox(2,1)[o] {\footnotesize {$k+1$}}}
\put(30,25){\circle * {0.25}}
\put(32,25){\circle * {0.25}}
\put(34,25){\circle * {0.25}}
\put(36,24.8){\line(1,0){5.1}}   
\put(36,25.2){\line(1,0){5.1}}
\put(36,25){\circle * {0.44}}
\put(35,23.2){\makebox(2,1)[o] {\footnotesize {$l-1$}}}
\put(37.5,24.494){\makebox(2,1)[o] {$<$}}
\put(41,25){\circle{0.5}}
\put(40,23.2){\makebox(2,1)[o] {\footnotesize {$l$}}}
\put(41.25,25){\line(1,0){4.8}}
\put(45.8,24.8){\makebox(0,0)[o] {\bf *}}
\put(45,23.2){\makebox(2,1)[o] {\footnotesize {$l+1$}}}
\put(0,16.5){\makebox(2,1)[l]{$D_l$}}
\put(5,17){\line(1,0){5}}   
\put(5,17){\circle * {0.44}} 
\put(4,15.2){\makebox(2,1)[o] {\footnotesize {$1$}}}
\put(10,17){\circle * {0.44}}  
\put(9,15.2){\makebox(2,1)[o] {\footnotesize {$2$}}}
\put(12,17){\circle * {0.25}}
\put(14,17){\circle * {0.25}}
\put(16,17){\circle * {0.25}}
\put(18,17){\circle * {0.44}}
\put(17,15.2){\makebox(2,1)[o] {\footnotesize {$k-1$}}}
\put(18,17){\line(1,0){5}}
\put(23,17){\line(1,0){5}}
\put(23,17){\circle * {0.44}}
\put(22,15.2){\makebox(2,1)[o] {\footnotesize {$k$}}}
\put(28,17){\circle * {0.44}}
\put(27,15.2){\makebox(2,1)[o] {\footnotesize {$k+1$}}}
\put(30,17){\circle * {0.25}} 
\put(32,17){\circle * {0.25}}
\put(34,17){\circle * {0.25}}
\put(36.1,17.3){\line(2,1){4}}
\put(36.1,16.7){\line(2,-1){4}}
\put(36,17){\circle{0.5}}
\put(34,15.2){\makebox(2,1)[o] {\footnotesize {$l-2$}}}
\put(39.1,20.2){\makebox(2,1)[o] {\footnotesize {$l-1$}}}
\put(40.1,19.2){\circle * {0.44}}
\put(40.1,14.8){\circle * {0.44}}
\put(39.1,13){\makebox(2,1)[o] {\footnotesize {$l$}}}
\put(40.1,14.8){\line(1,0){5}}
\put(44.9,14.6){\makebox(0,0)[o] {\bf *}}
\put(44.1,13){\makebox(2,1)[o] {\footnotesize {$l+1$}}}
\put(0,6.5){\makebox(2,1)[l]{$E_6$}}
\put(5,7){\line(1,0){9.75}}
\put(5,7){\circle * {0.44}}   
\put(4,5.2){\makebox(2,1)[o] {\footnotesize {$1$}}}
\put(10,7){\circle * {0.44}}
\put(9,5.2){\makebox(2,1)[o] {\footnotesize {$3$}}}
\put(15,7){\circle{0.5}}
\put(15.25,7){\line(1,0){14.75}}
\put(15,7.25){\line(0,1){4.65}}
\put(15,11.75){\circle * {0.44}}
\put(15,11.5){\makebox(2,1) [o] {\footnotesize {$2$}}}
\put(14,5.2){\makebox(2,1)[o] {\footnotesize {$4$}}}
\put(19,5.2){\makebox(2,1)[o] {\footnotesize {$5$}}}
\put(24,5.2){\makebox(2,1)[o] {\footnotesize {$6$}}}
\put(29.8,6.8){\makebox(0,0)[o] {\bf *}}
\put(29,5.2){\makebox(2,1)[o] {\footnotesize {$7$}}}
\put(20,7){\circle * {0.44}}
\put(25,7){\circle * {0.44}}
\put(0,-3.5){\makebox(2,1)[l]{$E_7$}}
\put(5,-3){\line(1,0){9.75}}
\put(5,-3){\circle * {0.44}}
\put(4,-5.2){\makebox(2,1)[o] {\footnotesize {$1$}}} 
\put(10,-3){\circle * {0.44}}
\put(9,-5.2){\makebox(2,1)[o] {\footnotesize {$3$}}} 
\put(15,-3){\circle{0.5}}  
\put(15.25,-3){\line(1,0){19.75}}
\put(15,-2.75){\line(0,1){4.7}}
\put(15,1.75){\circle * {0.44}}
\put(15,1.5){\makebox(2,1) [o] {\footnotesize {$2$}}}
\put(14,-5.2){\makebox(2,1)[o] {\footnotesize {$4$}}}
\put(19,-5.2){\makebox(2,1)[o] {\footnotesize {$5$}}}
\put(24,-5.2){\makebox(2,1)[o] {\footnotesize {$6$}}}
\put(29,-5.2){\makebox(2,1)[o] {\footnotesize {$7$}}}
\put(34.8,-3.2){\makebox(0,0)[o] {\bf *}}
\put(34,-5.2){\makebox(2,1)[o] {\footnotesize {$8$}}}
\put(20,-3){\circle * {0.44}}
\put(25,-3){\circle * {0.44}} 
\put(30,-3){\circle * {0.44}}
\put(0,-13.5){\makebox(2,1)[l]{$E_8$}}
\put(5,-13){\line(1,0){9.75}}
\put(5,-13){\circle * {0.44}}
\put(4,-15.2){\makebox(2,1)[o] {\footnotesize {$1$}}} 
\put(10,-13){\circle * {0.44}}
\put(9,-15.2){\makebox(2,1)[o] {\footnotesize {$3$}}}
\put(15,-13){\circle{0.5}}
\put(15.25,-13){\line(1,0){24.75}}
\put(15,-12.75){\line(0,1){4.7}}
\put(15,-8.25){\circle * {0.44}}
\put(15,-8.5){\makebox(2,1) [o] {\footnotesize {$2$}}}
\put(14,-15.2){\makebox(2,1)[o] {\footnotesize {$4$}}}
\put(19,-15.2){\makebox(2,1)[o] {\footnotesize {$5$}}}
\put(24,-15.2){\makebox(2,1)[o] {\footnotesize {$6$}}}
\put(29,-15.2){\makebox(2,1)[o] {\footnotesize {$7$}}}
\put(39,-15.2){\makebox(2,1)[o] {\footnotesize {$9$}}}
\put(39.8,-13.2){\makebox(0,0)[o] {\bf *}}
\put(34,-15.2){\makebox(2,1)[o] {\footnotesize {$8$}}}
\put(20,-13){\circle * {0.44}}  
\put(25,-13){\circle * {0.44}}
\put(30,-13){\circle * {0.44}}  
\put(35,-13){\circle * {0.44}}
\put(0,-21.5){\makebox(2,1)[l]{$F_4$}}
\put(10,-21){\line(1,0){4.75}}
\put(10,-21){\circle * {0.44}}
\put(15,-21){\circle{0.5}}
\put(9,-23.2){\makebox(2,1)[o] {\footnotesize {$1$}}} 
\put(14,-23.2){\makebox(2,1)[o] {\footnotesize {$2$}}}
\put(19,-23.2){\makebox(2,1)[o] {\footnotesize {$3$}}}
\put(24,-23.2){\makebox(2,1)[o] {\footnotesize {$4$}}}
\put(29,-23.2){\makebox(2,1)[o] {\footnotesize {$5$}}}
\put(20,-21){\line(1,0){10}}
\put(20,-21){\circle * {0.44}}
\put(25,-21){\circle * {0.44}}
\put(29.8,-21.2){\makebox(0,0)[o] {\bf *}}
\put(15,-21.2){\line(1,0){5}}
\put(15,-20.8){\line(1,0){5}}
\put(17.2,-21.507){\makebox(2,1)[o] {$>$}}
\put(0,-26.5){\makebox(2,1)[l]{$G_2$}}
\put(15,-26){\line(1,0) {4.75}}
\put(20.25,-26){\line(1,0) {4.75}}
\put(15,-26.2){\line(1,0) {5}}
\put(15,-25.8){\line(1,0) {5}}
\put(15,-26){\circle * {0.44}}
\put(20,-26){\circle {0.5}}
\put(16.2,-26.507){\makebox(2,1)[o] {$<$}}
\put(24.8,-26.2){\makebox(0,0)[o] {\bf *}}
\put(14,-28.2){\makebox(2,1)[o] {\footnotesize {$1$}}}
\put(19,-28.2){\makebox(2,1)[o] {\footnotesize {$2$}}}
\put(24,-28.2){\makebox(2,1)[o] {\footnotesize {$3$}}}
\put(23.9,-35){\makebox(2,1)[o] {Table 1}}
\end{picture}
\end{figure}
\noindent omitted)\,
consists of $\ 1,2$\  or $\ 3$\ branches of $\ A_r$\ type for
some $r$. Another observation is that the number
$$
\frac12 (\al_k,\al_k)
$$
is an integer for our choice of $k$.\par
We construct a group
$$
\tw=\tw^{(k)}(R)
$$
acting in
$$
\tv=V\oplus \bold R
$$
generated by the transformations
\begin{equation}
x=(\bx,x_{l+1})\mapsto (w(\bx)+\sum_{j=1}^l m_j\av_j,
\ x_{l+1}),\quad w\in W,\ m_j\in \bold Z,\tag{1.7a}
\end{equation}
and
\begin{equation}
x=(\bx,x_{l+1})\mapsto (\bx+\w_k,\ x_{l+1}-1).\tag{1.7b}
\end{equation}
\noindent {\bf {Definition.}}\ \  $\cal A=\cal A^{(k)}(R)$\ \ is 
the ring of all $\ \tw$-invariant Fourier polynomials of \newline
$\ x_1,
\cdots,x_l,\frac1{f} {x_{l+1}}$ \ that are bounded in the limit
\begin{equation}
\bx=\bx^{0}-i\ \w_k\tau,\quad x_{l+1}=x_{l+1}^{0}+i\ \tau,\quad
\tau\to +\infty,\tag{1.8}
\end{equation}
for any $\ x^{0}=(\bx^{0},x_{l+1}^{0})$,\ here $f$ is  
the determinant of the Cartan 
matrix $\ A$ \ of the root system $\ R$, see Table 2. \par
We put
\begin{equation}
d_j=(\w_j,\w_k),\quad j=1,\dots,l,\tag{1.9}
\end{equation}
these are certain positive rational numbers that can be found in
Table 2. All the numbers $\ f\cdot d_j$\ are integers. Indeed, they 
are the elements of the $\ k$-th column of the matrix $\ A^{-1}$\ 
times $\ \frac12 (\al_k, \al_k)$.\par
\def\qq{&\qquad} 
$$
\matrix
R\qq \ d_1,\dots,d_l\qq f\qq d_k\\ \\ \\
A_l\qq d_i=\cases\frac{i(l-k+1)}{l+1},\quad 1\le i\le k\\
              \frac{k(l-i+1)}{l+1},\quad k+1\le i\le l\endcases
   \qq l+1\qq \frac {k(l-k+1)}{l+1}\\
B_l\qq d_i=\cases i,\quad 1\le i\le l-1\\
              \frac{l-1}{2},\quad i=l\endcases
   \qq 2\qq l-1\\
C_l\qq d_i=i\qq 2\qq l\\
D_l\qq d_i=\cases i,\quad 1\le i\le l-2\\
              \frac{l-2}{2},\quad i=l-1,\ l\endcases
\qq 4\qq l-2\\ 
E_6\qq 2,\ 3,\ 4,\ 6,\ 4,\ 2\qq 3\qq 6\\
E_7\qq 4,\ 6,\ 8,\ 12,\ 9,\ 6,\ 3\qq 2\qq 12\\
E_8 \qq 10,\ 15,\ 20,\ 30,\ 24,\ 18,\ 12,\ 6\qq 1 \qq 30\\ 
F_4\qq 3,\ 6,\ 4,\ 2\qq 1\qq 6\\   
G_2\qq 3,\ 6\qq 1\qq 6
\endmatrix
$$
\vskip 0.5cm
$$
\text {Table 2}
$$
\vskip 0.5cm
\vskip 0.5cm
\begin{Lem} The Fourier polynomials
\begin{equation*}
\begin{aligned}
&\ty_j(x)=e^{2\pi i d_j x_{l+1}}\ y_j(\bx),\quad j=1,\dots,l,\\
&\ty_{l+1}(x)=e^{2\pi i x_{l+1}},   
\end{aligned}\tag{1.10}
\end{equation*}
are $\tw$-invariant.
\end{Lem}
\begin{pf} We show that all $\ \ty_1,\dots,\ty_l$\ are
$\tw$-invariant (invariance of $\ty_{l+1}$ is obvious). It suffices
to prove that
\begin{equation}
y_j(\bx+\w_k)=e^{2\pi i d_j}\ y_j(\bx).\tag{1.11}
\end{equation}
We can represent $\ y_j(\bx)$\ as
\begin{equation}
y_j(\bx)=n_j^{-1}\,\sum_{w\in W} e^{2\pi i (w(\w_j),\bx)},\tag{1.12}
\end{equation}
where $\, n_j=n_f$\, for $\,f=e^{2\pi i (\w_j,\bx)}$.\
According to [B, VI, \S 1.6, Prop. 18] for any $\ w\in W$
\begin{equation}
w(\w_j)=\w_j-\sum_{i=1}^l m_i\al_i\tag{1.13}
\end{equation}
for some non-negative integers $\ m_1,\dots,m_l$.\ \ So
$$
(w(\w_j),\w_k)=(\w_j,\w_k)-\sum_{i=1}^l m_i (\al_i,\w_k)
=d_j-m_{w},
$$
for an integer
\begin{equation}
m_{w}=\frac12 m_k\ (\al_k,\al_k),\tag{1.14}
\end{equation}
this leads to (1.11), and  we proved the lemma.
\end{pf}
Let us prove now boundedness of the functions $\ \ty_1,\dots,
\ty_l$\ in the limit (1.8).
\begin{Lem} In the limit 
\begin{equation}
\bx=\bx^{0}-i\ \w_k\tau,\quad \tau\to +\infty\tag{1.15}
\end{equation}
the functions $\ y_1(\bx),\dots,y_l(\bx)$\ have the
expansion
\begin{equation}
y_j(\bx)=e^{2\pi d_j\tau}[y_j^{0}(\bx^{0})+\cal O(e^{-2\pi \tau})],
\quad j=1,\dots,l,\tag{1.16}
\end{equation}
where
\begin{equation}
y_j^{0}(\bx^0)=n_j^{-1}\,\sum\Sb w\in W\\ (w(\w_j)-\w_j,\w_k)=0\endSb 
e^{2\pi i (w(\w_j),\bx^0)}.\tag{1.17}
\end{equation}
\end{Lem}
\begin{pf} From the representation (1.13) we see that 
the exponential \newline $\ \exp[2\pi i (w(\w_j),\bx)]$\
in the limit (1.15) behaves as
$$
e^{2\pi(d_j-m_w)\tau}\ e^{2\pi i (w(\w_j),\bx^{0})},
$$
where the non-negative integer $m_w$ is defined in (1.14).
Thus the leading contribution in the asymptotic behaviour of
the sum (1.12) comes from those $\ w\in W$\ satisfying $\ m_k=0$.\
Lemma is proved.\end{pf}
\begin{Cor} The functions $\ \ty_1(x),\dots,
\ty_{l+1}(x)$\ belong to $\cal A$.
\end{Cor}
\begin{pf} From (1.16) it follows that 
\begin{equation*}
\begin{aligned}
&\ty_j(x)\to \ty_j^0(x^0)=e^{2\pi i d_j x_{l+1}^0}\ y_j^0(\bx^0),\quad 
j=1,\dots,l,\\
&\ty_{l+1}\to 0
\end{aligned}
\end{equation*}
in the limit (1.8), where the functions $\ y^0_j(\bx^0)$\ are defined 
in (1.17).
Corollary is proved.\end{pf}
The main result of this section is
\begin{Thm} The ring $\cal A$ is isomorphic to
the ring of polynomials of $\ \ty_1,\dots,\ty_{l+1}$.
\end{Thm}
\begin{pf} We will show that any element $f(x)$ of the ring 
$\cal A$ 
can be represented as a polynomial of  $\ \ty_1,\dots,\ty_{l+1}$.\
This will be enough since the functions  $\ \ty_1,\dots,\ty_{l+1}$
\ are algebraically independent.\par
From the invariance w.r.t. $\ \tw$\ it easily follows (using theorem [B])
that any $f(x)$ can be represented as a polynomial of  $\ 
\ty_1(x),\dots,\ty_l(x),\ty_{l+1}(x),\ty_{l+1}^{-1}$.\
We need to show that in $f(x)$ there are no negative powers of
$$
\ty_{l+1}(x)=e^{2\pi i x_{l+1}}.
$$
Let's assume 
$$
f(x)=\sum_{n\ge -N} \ty_{l+1}^n\ P_n(\ty_1(x),\dots,\ty_{l}(x)),
$$
and the polynomial 
$\ P_{-N}(\ty_1(x),\dots,\ty_{l}(x))$\
does not vanish identically for certain positive integer $N$. 
From Corollary 1.1
we obtain that in the limit (1.8) the function $f(x)$ behaves as
$$
f(x)=e^{2\pi N\tau-2\pi i N x_{l+1}^0} 
[P_{-N}(\ty_1^{0}(x^0),\dots,\ty_{l}^{0}(x^0))+ \cal O(e^{-2\pi \tau})],
$$
where 
$$
\ty_j^{0}(x^0)=e^{2\pi i\ d_jx_{l+1}^{0}}\ y_j^{0}(\bx^0),\quad j=1,
\cdots,l,
$$
and the functions $\ y_j^{0}(\bx^0)$\ are defined in (1.17).\par
To obtain a function bounded for $\ \tau\to +\infty$\ it is necessary
to have 
$$
P_{-N}(\ty_1^{0}(x^{0}),\dots,\ty_{l}^{0}(x^{0}))\equiv 0
$$
for any $x^{0}=(\bx^0,x_{l+1}^0)$.\ 
We show now that this is impossible 
due to algebraic independence of the functions $\ \ty_1^{0},\dots,
\ty_l^{0}$.\
It is sufficient to prove algebraic independence of the
functions $\ \zy_1(\bx),\dots,\zy_l(\bx)$.
\begin{MainLem} The Fourier polynomials $\ \zy_1(\bx),
\cdots,\zy_l(\bx)$\ are algebraically independent.
\end{MainLem}
We will prove that these functions are functionally independent,
i.e., that the Jacobian
$$
\det(\frac{\pal \zy_j(\bx)}{\pal x_i})
$$
does not vanish identically.
At this end we derive explicit formulae for these functions and then
prove nonvanishing of the Jacobian.\par
Consider 
$$ 
R_k =R^{(1)}\cup R^{(2)}\cup\dots,
$$
here any subsystem $\ R^{(1)},\dots$ is a root system of the type
$\ A_r$\ for some $r$ (see Table 1).\par
Let $\ \w_i$\ be a fundamental weight orthogonal to 
$\ R\setminus R^{(1)}$.
\begin{Lem} Let for some $\ w\in W(R)$
$$
w(\w_i)=\w_i-\sum_{m=1}^l c_{m}\al_m
$$
such that $\ c_k=0$,\ then $\ c_m\ne 0$\ only if $\ \al_m\in R^{(1)}$.
\end{Lem}
\begin{pf} In $\ W(R_k\setminus R^{(1)})$\ there exists $\ w_0$\ 
such that it maps all positive roots of $\ R_k\setminus R^{(1)}$\
into negative roots. Clearly, $w_0$\ preserves all $\ \al_m\in R^{(1)}$,\
and $\ w_0(\w_i)=\w_i$.\ We represent 
$$
w(\w_i)=\w_i -\al^{(1)}-\al^{(2)}-\cdots,
$$
where $\ \al^{(i)}$'s\ are sum of some positive  roots 
of $\ R^{(i)}$.\
Then
$$
w_0w(\w_i)=\w_i-\al^{(1)}-w_0(\al^{(2)}+\cdots)
=\w_i-\al^{(1)}+\sum_{\al_m\in R_k\setminus R^{(1)}} \tilde c_m\al_m
$$
for some non-negative integers $\ \tilde c_m$,\ and not all of these
integers vanish if there exists certain $\ \al_m\notin R^{(1)}$\
such that $\ c_m\ne 0$.\ This
contradicts to negativity of $\ w_0w(\w_i)-\w_i$.\ Lemma
is proved.
\end{pf}
\par
A similar statement holds true for other components $\ R^{(2)},\cdots$
(if any) of $\ R_k$.\par 
\begin{Lem} If for some $\ w\in W(R)$
$$
w(\w_k)=\w_k-\sum_{m\ne k} c_m\al_m,
$$
then all $\ c_m=0$.\end{Lem}
\begin{pf} There exists $\ w_0\in W(R_k)$\ such that it maps any positive
roots of $\ R_k$\ into negative ones, and preserves $\w_k$.\ So
$$
w_0w(\w_k)=\w_k+\text {sum of some positive roots},
$$
which leads to the result of the lemma.
\end{pf}
\begin{Lem} Under the assumption of Lemma 1.3 there
exists 
$\ \tilde w\in W(R^{(1)})$\ such that 
$$
\tilde w(\w_i)=w(\w_i)=\w_i-\sum_{\al_m\in R^{(1)}} c_m\al_m.
$$
\end{Lem}
\begin{pf} We 
will use induction on the length of $w$.\ If the length of 
$w$ equals one, then the lemma holds true obviously.
We now assume that the lemma holds true when the length of $w$
is less than $p$.  
Let $\ w$\ has the reduced expression $\ \sg_{i_1}\cdots\sg_{i_p}$,\
then it follows from Lemma 1.3 that
\begin{equation}
\sg_{i_1}\cdots\sg_{i_p}(\w_i)=\w_i-\sum_{\al_m\in R^{(1)}} c_m\al_m.
\tag{1.18}
\end{equation}
If all $\ c_m=0$,\ then we can put $\ \tilde w=1$,\ otherwise we 
rewrite (1.18) in the form
$$
\w_i=\sg_{i_p}\sg_{i_{p-1}}\cdots\sg_{i_1}(\w_i)-
\sum_{\al_m\in R^{(1)}}
c_m\sg_{i_p}\sg_{i_{p-1}}\cdots\sg_{i_1}(\al_m).
$$
We put 
$$
\sg_{i_p}\sg_{i_{p-1}}\cdots\sg_{i_1}(\w_i)=\w_i-\sum_{m=1}^l  b_m\al_m
$$
for some non-negative integres $\ b_1,\dots,b_l$.\ We claim now that 
there exists a root $\ \al_{m_1}\in R^{(1)}$\ such that $\
c_{m_1}\ne 0$ in (1.18) and
$$
w^{-1}(\al_{m_1})=\sg_{i_p}\sg_{i_{p-1}}\cdots\sg_{i_1}(\al_{m_1})
$$
is a negative root. Indeed, otherwise the root
$$
\sum_{m=1}^l b_m\al_m+\sum_{\al_m\in R^{(1)}} c_m 
\sg_{i_p}\sg_{i_{p-1}}\cdots\sg_{i_1}\al_m
$$
could not be equal to zero since $c_m$'s are non-negative
integers. We use now the 
following proposition:
\begin{PropH} Let 
$\al_{j_1},\dots,\al_{j_t}$\
be some simple roots of $R$ (not necessarily distinct). If
$$
\sg_{j_1}\cdots\sg_{j_{t-1}}(\al_{j_t})
$$
is a negative root, then for some $\ 1\le s\le t-1$\ we have
$$
\sg_{j_1}\cdots\sg_{j_{t-1}}=\sg_{j_1}\cdots\hat{\sg}_{j_s}
\cdots\sg_{j_{t-1}}\sg_{j_t},
$$
where the hat above $\sg_{j_s}$ means that this factor is omitted
in the product.
\end{PropH}
According to this statement we can represent
$$
w^{-1}=\sg_{i_p}\cdots\hat {\sg}_{i_s}\cdots\sg_{i_1}\sg_{m_1}
$$
for some $\ 1\le s\le p$.\ \
We can now rewrite (1.18) as 
$$
w'(\w_i)=\sg_{m_1}(\w_i)-\sum_{\al_m\in R^{(1)}} c_m\sg_{m_1}(\al_m)
=\w_i-\sum_{\al_m\in R^{(1)}} c_m'\al_m
$$
for some new non-negative integers $\ c_m'$,\ and for
$$
w'=\sg_{i_1}\cdots\hat{\sg}_{i_s}\cdots\sg_{i_p}.
$$
The length of $\ w'$\ is less than the length $p$ of $w$. Using the 
induction assumption we complete the proof of the lemma.
\end{pf}
\begin{Cor}
1)\  \ Let $R^{(s)}$ be any branch of $R_k$, then for any 
$\ \al_j\in R^{(s)}$\ we have
\begin{equation}
\zy_j(\bx)=m_j^{-1}\,\sum_{w\in W(R^{(s)})} e^{2\pi 
i(w(\w_j),\bx)};\tag{1.19a} \end{equation}
\newline
2) 
\begin{equation}
\zy_k(\bx)=e^{2\pi i(\w_k,\bx)}=e^{2\pi i x_k},\tag{1.19b}
\end{equation}
where $\,m_j=\# \{w\in W(R^{(s)})\,|\,w(\w_j)=\w_j\}$.
\end{Cor}
\noindent
{\it {Proof of the main lemma.}}\ \ 
We proceed now to the proof of algebraic independence of the functions
$\zy_1(\bx),\dots, 
\zy_l(\bx)$\ by analyzing the formulae (1.19)
for all the cases of root systems.
Let's define $\ z_i,\ i=1,\dots,l$\ as in (1.5), and denote the
$j$-th order elementary symmmetric polynomial of $n$ variables
$\ u_1,\dots,u_n$ \ by $\ s_j(u_1,\dots,u_n)$ \ with $\ 
s_0(u_1,\dots,u_n)=1$.  \newline
1)\ \ For the root system of type $A_l$, 
from Example 1.1 and (1.19) we have
\begin{equation*}
\begin{aligned}
&\zy_i(\bx)=v_i+v_{i-1}v_k,\quad 1\le i\le k-1\\
&\zy_k(\bx)=v_{k-1}v_{k},\\
&\zy_{k+1}(\bx)=v_{k+1}v_{k-1}v_k+\frac1{v_l},\\
&\zy_{k+j}(\bx)=v_{k+j}v_{k-1}v_k+\frac{v_{k+j-1}}{v_{l}},\quad 2\le
j\le l-k,
\end{aligned}\tag{1.20}
\end{equation*}
where
\begin{equation*}
\begin{aligned}
&v_i=s_i(e^{2\pi i z_1},\dots,e^{2\pi i z_{k-1}}),
\quad 1\le i\le k-1,\\
&v_{k}=e^{2\pi i z_k},\\
&v_{k+j}=s_j(e^{2\pi i z_{k+1}},\dots,e^{2\pi i z_l}),\quad
1\le j\le l-k.
\end{aligned} \tag{1.21}
\end{equation*}
Clearly the variables $\ v_1,\dots,v_l$\
are algebraically independent.
We have
\begin{equation}
\det(\frac{\pal \zy_i}{\pal v_j})=(-1)^{k-1}v_{k-1}^{l-k} v_k^{l-1}
+\text {lower order terms of}\ \ v_k.\tag{1.22}
\end{equation}
So the Jacobian $\ \det(\frac{\pal \zy_i}{\pal x_j})$\ does not vanish 
identically.\newline
2)\ \ For the root system of type $B_l$\ and $C_l$, \
we have $\ k=l-1$\ and $\ k=l$\ respectively, all
the formulae in (1.20)-(1.22) hold true if we replace $k$
by $l-1$\ and $l$\ respectively.
So in these cases the Jacobian does not vanish identically neither.
\newline
3)\ \ For the root system of type $D_l,\ k=l-2$,\ we have
\begin{equation*}
\begin{aligned}
&\zy_i(\bx)=v_i+v_{i-1}v_{l-2},\quad 1\le i\le l-3\\
&\zy_{l-2}(\bx)=v_{l-3}v_{l-2},\\
&\zy_{l-1}(\bx)=v_{l-3}v_{l-2}v_{l-1}+\frac{1}{v_{l-1}},\\
&\zy_l(\bx)=v_{l-3}v_{l-2}v_{l-1}v_l+\frac{1}{v_{l-1}v_l},
\end{aligned}\tag{1.23}
\end{equation*}
where $\ v_i,\ 1\le i\le l-2$ \ are defined in (1.21)
with $k=l-2$,\ and $\ v_{l-1}=e^{2\pi i z_{l-1}},\
v_{l}=e^{2\pi i z_l}$.\
We have
$$
\det(\frac{\pal \zy_i}{\pal v_j})=(-1)^{l+1}
v_{l-1}v_{l-3}^2v_{l-2}^{l-1}
+\text {lower order terms of}\ \ v_{l-2}.
$$
Since the functions $\ v_1,\dots,v_l$\ are algebraically
independent, the
Jacobian $\ \det(\frac{\pal \zy_i}{\pal x_j})$\ does not vanish
identically.\newline
4)\ \ For the root system of type $E_l,\ k=4$,\ \ let's define
$$
\beta_1=x_l,\ \beta_i=x_{l-i+1}-x_{l-i+2},\quad 2\le i\le l,
$$
and
\begin{equation*}
\begin{aligned}
&v_i=s_i(e^{2\pi i\beta_1},\dots,e^{2\pi i\beta_{l-4}}),
\quad 1\le i\le l-4,\\
&v_{l-3}=e^{2\pi i \beta_{l-3}},\\
&v_{l-2}=e^{2\pi i (\beta_{l-2}+\beta_{l-1}+\beta_l)},\\
&v_{l-1}=e^{2\pi i(\beta_{l-1}+\beta_l)},\\
&v_l=e^{2\pi i \beta_l}.
\end{aligned}
\end{equation*}
Then we have
\begin{equation*}
\begin{aligned}
&\zy_1=v_{l-4} v_{l-3}v_{l-2}+\frac1{v_{l-1}}
+\frac{v_{l-1}}{v_{l-2}},\\
&\zy_2=v_{l-4} v_{l-3} \frac {v_{l-2}}{v_{l}}+\frac{v_l}{v_{l-2}},\\
&\zy_3=v_{l-4} v_{l-3} \frac{v_{l-2}}{v_{l-1}}+v_{l-4}v_{l-3}v_{l-1}+
\frac1{v_{l-2}},\\
&\zy_4=v_{l-4}v_{l-3},\\
&\zy_{l-i+1}=v_i+v_{i-1}v_{l-3},~1\leq i \leq l-4
\end{aligned}\tag{1.24}
\end{equation*}
and
$$
\det(\frac{\pal\zy_i}{\pal v_j})=\epsilon \left( 
1-\frac{v_{l-2}}{v_{l-1}^2} \right) \frac{v_{l-2}
(v_{l-4})^{3}}{v_l^2}v_{l-3}^{l-1}+\text {lower order terms of}\ \
v_{l-3},
$$
where $\epsilon=-1$\ or $\ 1$,\
which shows that the
Jacobian $\ \det(\frac{\pal \zy_i}{\pal x_j})$\ does not vanish
identically.\newline
5)\ \ For the root system of type $F_4,\ k=2$,\ define $\beta_i$
as for the $E_l$ case with $l=4$, we have
\begin{equation*}
\begin{aligned}
&\zy_1=v_2v_3v_4+\frac1{v_4},\\
&\zy_2=v_2v_3,\\
&\zy_3=v_2+v_1v_3,\\
&\zy_4=v_1+v_3,
\end{aligned}\tag{1.25}
\end{equation*}
where $\ v_j=s_j(e^{2\pi i \beta_1},e^{2\pi i \beta_2}),\ j=1,2,\ \
v_3=e^{2\pi i \beta_3},\ v_4=e^{2\pi i\beta_4}$. Then
$$
\det(\frac{\pal\zy_i}{\pal v_j})=v_2 v_3^3
+\text {lower order terms of}\ \ v_{3}.
$$
Since the functions $\ v_1,\dots,v_4$\ are algebraically
independent, the
Jacobian $\ \det(\frac{\pal \zy_i}{\pal x_j})$\ does not vanish
identically.\newline
6)\ \ For the root system of type $G_2,\ k=2$, we have
\begin{equation}
\zy_1=e^{2\pi i x_1}+e^{2\pi i(x_1-x_2)},\quad
\zy_2=e^{2\pi i x_2}.\tag{1.26}
\end{equation}
so $\zy_1$ and $\zy_2$ are algebraically independent.\par
We thus proved the Main lemma and also Theorem 1.1.
\end{pf}
\begin{Cor} The function $\deg$ defined as
\begin{equation*}
\begin{aligned}
&\deg \ty_j=d_j,\quad j=1,\dots,l,\\
&\deg \ty_{l+1}=1\end{aligned}
\end{equation*}
determines on $\cal A$ a structure of graded polynomial ring.
\end{Cor}
\par
We end this section with an important observation about the
numbers $\ d_1,\dots,d_l$.\ Let us again consider the
components of the Dynkin graph of $\ R_k=R\setminus \al_k$.\ \
By $R^{(1)}$ we denote the component that touches the added
vertex (the asterisk) on Table 1. We put
$$
\hat{R}^{(1)}=R^{(1)}\cup \{\al_k\}\cup \{*\}.
$$
This is again an $A_r$-type diagram. By $\ R^{(2)},\cdots$
we denote other components of $R_k$.\ We put also
$$
d_{l+1}=0.
$$
On any of the diagram $\ \hat {R}^{(1)},\ R^{(2)},\cdots$
there is an involution
$$
\al_i\mapsto \al_{i^*}
$$
corresponding to the reflection of the component w.r.t. the
center.
\begin{Lem} The numbers $\ d_1,\dots,d_{l+1}$
satisfy the duality relation
\begin{equation}
d_i+d_{i^*}=d_k.\tag{1.27}
\end{equation}
\end{Lem}
\begin{pf} By obvious inspection of Table 2.\end{pf}
\par
\section{Differential geometry of the orbit spaces 
of extended affine Weyl groups}
Let $\ \cal M=\cal {M}(R,k)=\text{Spec}\cal A$.\  We call it {\it {orbit
space}} of the extended affine Weyl group $\ \tw$.\ According
to Theorem 1.1 and Corollary 1.3 this is a graded affine algebraic 
variety of the dimension $\ n=l+1$.\ The functions $\ \ty_1(x),
\cdots,\ty_{l+1}(x)$\ serve as global coordinates on $\cal M$.\ We 
will however use the local coordinates $\ y^1=\ty_1(x),
\cdots,y^l=\ty_l(x)$\ and 
$\ y^{l+1}=\log \ty_{l+1}(x)=2\pi i x_{l+1}$.\ The last coordinate is
multivalued on $\cal M$.\ In other words, it lives on a covering
$\widetilde{\cal M}$\ of $\ \cal M\setminus\{\ty_{l+1}=0\}$.\par
The projection map
$$
P:\ \widetilde V\to \widetilde{\cal M}
$$
is given by the formulae
\begin{equation}
(x_1,\dots,x_{l+1})\mapsto (\ty_1(x),\dots,\ty_l(x),2\pi i x_{l+1})
=(y^1,\dots,y^{l+1}).\tag{2.1}
\end{equation}
For the Jacobian of the projection map we have a formula
\begin{equation*}
\begin{aligned}
&\det(\frac{\pal y^j}{\pal x_p})=2\pi i\ e^{2\pi i(d_1+\cdots+d_l) x_{l+1}}
\det(\frac{\pal y_j(\bx)}{\pal x_p})\\
&=c\ \exp[2\pi i(d_1+\cdots+d_{l})x_{l+1}-\sum_{\al\in \Phi^+}
\pi i(\al,\bx)]\prod_{\beta\in\Phi^+}(e^{2\pi i(\beta,\bx)}-1),
\end{aligned}\tag{2.2}
\end{equation*}
where $\ \Phi^+$\ is the set of all positive roots and $c$ is
a non-zero constant [B, page 185, 228]. So the projection map is
a local diffeomorphism outside the hyperplanes 
\begin{equation}
\{(\bx,x_{l+1})|(\beta,\bx)=m\in\bold Z,\ x_{l+1}=\text {arbitrary}\},
\quad \beta\in \Phi^+.\tag{2.3}
\end{equation}
Recall that the hyperplanes $\ \{\bx |(\beta,\bx)=m\in\bold Z\}$\
are the mirrors of the affine Weyl group.\par
In this section, we will introduce a structure of Frobenius
manifold on $\cal M$\ (see definition below, and also in [D]).
We define first an indefinite metric $\ (\ ,\ )^{\sptilde}$\ \ on 
\newline
$\widetilde V
=V\oplus \bold R$.\ The restriction of $\ (\ ,\ )^{\sptilde}$\ onto $V$ 
coincides
with the $W$-invariant Euclidean metric $\ (\ ,\ )$\ on $V$ times
$\ 4\pi^{2}$.\ The coordinate $x_{l+1}$ is orthogonal to $V$\
(so $\ V\oplus \bold R$\ is orthogonal direct sum). Finally we put
$$
(e_{l+1},e_{l+1})^{\sptilde} 
=-{4\pi^2} (\w_k,\w_k)=-{4\pi^2} d_k,
$$
where $e_{l+1}$ is the unit vector along the $x_{l+1}$ axis.\par
We introduce now a symmetric bilinear form on $\ \ts$\ taking
projection of $\ (\ ,\ )^{\sptilde}$.\ More explicitly (cf. [A1]) the 
bilinear form on $\ \ts$\ in the coordinates $\ y^1,\dots,y^{l+1}$\
is given by a $\ (l+1)\times (l+1)$\ matrix $\ (g^{ij})$\ of the
form
\begin{equation}
(dy^i,dy^j)^{\sptilde}
\equiv g^{ij} :=\sum_{a,b=1}^{l+1}\frac{\pal y^i}{\pal x^a}
\frac{\pal y^j}{\pal x^b}(dx^a,dx^b)^{\sptilde},\tag{2.4}
\end{equation}
here $\ x^a=x_a, \ 1\le a\le l+1$,\ and this notation will also be used
later.
\begin{Lem} The matrix entries $g^{ij}$ of (2.4) 
are weighted homogeneous polynomials in $\ \ty_1,\dots,\ty_{l+1}$\
of the degree
$$
\deg g^{ij}=\deg y^i+\deg y^j,
$$
(here $\deg y^{l+1}=d_{l+1}=0$).\ The matrix $\ (g^{ij})$\
does not degenerate outside the $P$-images of the hyperplanes (2.3)
\end{Lem}
\begin{pf} We have for $\ 1\le i,j\le l$\ 
\begin{equation}
g^{ij}=\frac{d_i d_j}{d_k} \ty_i \ty_j+\frac1{4\pi^2}\sum_{p,q=1}^l
\frac{\pal\ty_i}{\pal x_p}\frac{\pal\ty_j}{\pal x_q} (\w_p,\w_q).\tag{2.5a}
\end{equation}
This is a Fourier polynomial invariant w.r.t. $ \tw$.\ Clearly it is 
bounded on the limit (1.8). According to Theorem 1.1 this is a polynomial
in $\ \ty_1,\dots,\ty_{l+1}$.\ The homogeneity is obvious.\par
For $\ g^{j ,l+1}$\ the computation is even simpler. For $\ 1\le j\le l$\
we obtain
\begin{equation}
g^{j,l+1}=\frac{d_j}{d_k}y^j.\tag{2.5b}
\end{equation}
Finally,
\begin{equation}
g^{l+1,l+1}=\frac 1{d_k}.\tag{2.5c}
\end{equation}
Nondegeneracy of $\ (g^{ij})$ \ follows from (2.4) and from
the formula (2.2) for the Jacobian. Lemma is proved.
\end{pf}
\par
According to Lemma 2.1, the image of nonregular orbits (2.3) is an 
algebraic subvariety $\ \Sigma$\ in $\cal M$ given by the polynomial
equation
$$
\Sigma=\{y|\det(g^{ij}(y))=0\}.
$$
We will call $\ \Sigma$\ the {\it {discriminant}} of the extended
affine Weyl group. On $\ \cal M\setminus\Sigma$\ the matrix
$\ (g^{ij})$\ is invertible; the inverse matrix
$$
(g_{ij})=(g^{ij})^{-1}
$$
determines a metric\footnote
{
The word `metric' in this paper will denote a symmetric bilinear
nondegenerate quadratic
form on $\ \text {T}\cal M$.\ The metric
is called flat if by a\ local 
change\ of coordinates  it can be reduced
to a constant form.%
}
on $\ \cal M\setminus\Sigma$.\ \ Of course, this is a flat metric
since it is obtained from a constant metric on $\widetilde V$
by a change of coordinates (see formula (2.4)).\par
Let us now compute the coefficients of the correspondent Levi-Civita
connection $\bn$ for the metric $(\ ,\ )^{\sptilde}$ 
defined by (2.4). It is
convenient to consider the ``contravariant components'' of the
connection
\begin{equation}
\Gamma^{ij}_m=(dy^i,\bn_m dy^j)^{\sptilde}.\tag{2.6}
\end{equation}
They are related to the standard Christoffel coefficients by the 
formula 
$$
\Gamma^{ij}_m=-g^{is}\Gamma^j_{sm}.
$$
For the contravariant components we have the formula
\begin{equation}
\Gamma^{ij}_mdy^m=\frac{\pal y^i}{\pal x^p}\frac{\pal^2y^j}
{\pal x^q\pal x^r}(dx^p,dx^q)^{\sptilde}\ dx^r,\tag{2.7}
\end{equation}
here and henceforth summation over the repeated indices is assumed.
\begin{Lem} $\Gamma^{ij}_m$'s\ are weighted homogeneous
polynomials in $\ \ty_1,\dots,\ty_{l+1}$ \ of the degree 
$\ \deg y^i+\deg y^j-\deg y^m$.\end{Lem}
\begin{pf} From (2.2) and (2.7) we can
represent 
$$
\Gamma^{ij}_m=e^{2\pi i(d_i+d_j-d_m)x_{l+1}}\frac{P^{ij}_m(\bx)}{J(\bx)},
$$
where $\ P^{ij}_m$\ is certain Fourier polynomial in $\ x_1,\dots,x_l$,
$$
J(\bx)=e^{-\sum_{\al\in \Phi^+}
\pi i(\al,\bx})\prod_{\beta\in\Phi^+}(e^{2\pi i(\beta,\bx)}-1)
$$
is anti-invariant w.r.t the  Weyl 
group $\ W$,\ it has a simple zero on any mirror of the 
Weyl group, and it changes sign w.r.t. the reflection in the mirrors.
But $\ \Gamma^{ij}_m$\ must be invariant w.r.t. the Weyl
group (it is invariant even w.r.t. the extended affine Weyl group $\ \tw$).\
So $\ P^{ij}_m$\ must be anti-invariant. Hence it is divisible
by $J(\bx)$ [B]. It follows that $\ \Gamma^{ij}_m$
is a Fourier polynomial in $\ x_1,\dots,x_l,\frac1{f} x_{l+1}$,\
where $f$ is the determinant of the Cartan matrix of the root system
(see Table 2). Since it is 
invariant w.r.t. the extended group $\ \tw$\ and is bounded in the
the limit (1.8), it belongs to $\cal A$,\ we conclude from
Theorem 1.1 that it is a polynomial in $\ \ty_1,\dots,\ty_{l+1}$.\
The homogeniety property is then obvious. Lemma is proved. 
\end{pf}
\begin{Cor} The polynomials $\ g^{ij}(y)$\ and
$\ \Gamma^{ij}_m(y)$\ are at most linear in $y^k$.
\end{Cor}
\begin{pf} This follows from weighted homogeneity and from
the following important observation:
$$
d_k>d_j\quad\text {for any}\ \ j\ne k,
$$
(see Table 2). We need however to prove linearity in $y^k$ of
the components $\ g^{kk}$\ and $\ \Gamma^{kk}_{l+1}$.\ According
to (2.5a) we have
\begin{equation*}
\begin{aligned}
g^{kk}=&d_k\ty_k^2+\frac1{4\pi^2}\sum_{p,q=1}^l 
\frac{\pal\ty_k}{\pal x_p}\frac{\pal\ty_k}{\pal x_q} (\w_p,\w_q)\\
=&e^{4\pi i d_k x_{l+1}}(d_k y_k^2+\frac1{4\pi^2}\sum_{p,q=1}^l
\frac{\pal y_k(\bx)}{\pal x_p}\frac{\pal y_k(\bx)}{\pal x_q} (\w_p,\w_q)).
\end{aligned}
\end{equation*}
The second term in the bracket is a $W$-invariant Fourier polynomial
of the form 
\begin{equation*}
\begin{aligned}
g&=\frac1{4\pi^2}\sum_{p,q=1}^l
\frac{\pal y_k(\bx)}{\pal x_p}\frac{\pal y_k(\bx)}{\pal x_q} (\w_p,\w_q)\\
&=-n_k^{-2}\,\sum_{w,w'\in W}(w(\w_k),w'(\w_k)) e^{2\pi 
i(w(\w_k)+w'(\w_k),\bx)}. \end{aligned}
\end{equation*}
We use now the standard partial ordering of the weights [H, page 69]:
$$
\w\succ \w'
$$ 
iff 
$$
\w-\w'=\sum_{m=1}^l c_m\al_m
$$
for some non-negative integers $\ c_1,\dots,c_l$.\ In this case we will
also write
$$
e^{2\pi i(\w,\bx)}\succ e^{2\pi i(\w',\bx)}.
$$
The $W$-invariant Fourier polynomial $g$ has unique maximal term
$$
-(\w_k,\w_k)e^{2\pi i(2\w_k,\bx)}=-d_k e^{2\pi i(2\w_k,\bx)}.
$$
Because of this all the terms in the $W$-invariant Fourier polynomial
\begin{equation}
d_k y_k^2+g\tag{2.8}
\end{equation}
are strictly less than $\ e^{2\pi i(2\w_k,\bx)}$.\
Hence the representation of (2.8) as a polynomial in $\
y_1(\bx),\dots,y_l(\bx)$\ does not contain $\ y_k^2$.\
That means that $\ g^{kk}$\ is at most linear in $y^k$.\par
For $\Gamma^{kk}_{l+1}$\ we have
\begin{equation*}
\begin{aligned}
\Gamma^{kk}_{l+1}=&\frac{\pal x^\gamma}{\pal y^{l+1}}\frac{\pal y^k}{\pal 
x^p} \frac{\pal^2 y^k}{\pal x^q\pal x^\gamma} (d x^p,d x^q)^{\sptilde}\\
=& \frac{\pal y^k}{\pal x^p}\frac{\pal}{\pal y^{l+1}}(\frac{\pal y^k}{\pal 
x^q}) (d x^p,d x^q)^{\sptilde}\\
=&\frac12 \frac{\pal g^{kk}}{\pal y^{l+1}}.\end{aligned}
\end{equation*}
so it also depends at most linearly on $y^k$. Corollary is proved.
\end{pf}
\par
We define a new metric  on $\ \ts$\ putting
\begin{equation}
\eta^{ij}(y)=\frac{\pal g^{ij}}{\pal y^k}.\tag{2.9}
\end{equation}
Up to multiplication by a nonzero constant this metric does not 
depend on the choice of basic homogeneous $\tilde W$-invariant 
Fourier polynomials. Indeed, since
$$\deg y^k > \deg y^j \ \ \text {for any} \ \ j\neq k
$$
the ambiguity in the choice of the basis $y^1, \dots, y^{l+1}$
is of the form
\begin{equation*}
\begin{aligned}
y^k &\mapsto cy^k + f^k \left( y^1, \dots, \hat y^k, \dots, 
y^l, \exp y^{l+1}\right)\\
y^j &\mapsto f^j \left( y^1, \dots, \hat y^k, \dots, y^l, 
\exp y^{l+1}\right), ~j\neq k, l+1 \\
y^{l+1} & \mapsto y^{l+1},
\end{aligned}
\end{equation*}
where $c$ is a nonzero constant and the polynomials $f^j$ are 
weighted homogeneous of the degree $d_j$ resp. So the vector field 
$\partial / \partial y^k$ is invariant within a constant:
$${\partial \over \partial y^k} \mapsto c {\partial \over \pal y^k}.
$$\par
The same formulae prove that the matrix $\eta^{ij}(y)$ behaves like a 
(2,0)-tensor (i.e., a symmetric bilinear form on the cotangent 
bundle) w.r.t. the changes of homogeneous coordinates on the orbit 
space.
\begin{MainLem} The determinant of the matrix 
$\ (\eta^{ij})$\ is a nonzero constant.\end{MainLem}
To prove this lemma, we need the following lemmas:
\begin{Lem} Let $R$ be a root system of type
$A_l-B_l-C_l-D_l$, \ denote $\ R_k\equiv R\setminus \al_k
=R^{(1)}\cup R^{(2)}\cup\cdots$.\newline
1)\ \ If $\ \al_i$\ and $\ \al_j$\
belong to different components of $R_k$, then $\ \eta^{ij}=0$.\newline
2)\ \ The block $\ \eta^{(t)}=(\eta^{ij})|_{\al_i,\al_j\in R^{(t)}}$\ \ of 
the matrix $\ \eta=(\eta^{ij})$\ corresponding to any 
branch $\ R^{(t)}$\ has triangular form, i.e. it has 
all zero entries above or under the antidiagonal. The antidiagonal
elements  of $\ \eta^{(t)}$\ consist of the constant
numbers $\ \eta^{ii^*}$\ for $\ \al_i\in R^{(t)}$.\newline
3)\ \ $\eta^{i(l+1)}=\eta^{(l+1)i}=\frac1{d_k}\delta_{ik}$.
\end{Lem} 
\begin{pf}  As we did in the proof of
Corollary 2.1, we use the standard partial ordering of the weights of 
$R$.\ When $\ 1\le i,j\le l$,\ we know from (2.5a) that
$\ g^{ij}=e^{2\pi i(d_i+d_j)x_{l+1}}\ h(\bx)$,\ where
$$
h(\bx)=\frac{d_i d_j}{d_k} y_i(\bx) y_j(\bx)
-(n_i\,n_j)^{-1}\,\sum_{w,w'\in W}(w(\w_i),w'(\w_j)) e^{2\pi 
i(w(\w_i)+w'(\w_j),\bx)}. $$
All the terms in the Fourier polynomial $\ h(\bx)$\
are strictly less than $\ e^{2\pi i (\w_i+\w_j,\bx)}$\
except the term
$\ c\ e^{2\pi i (\w_i+\w_j,\bx)},
$\
where $c$ is certain constant. 
So if $h({\bx})$ as a polynomial in $y_1, \dots, y_l$ contains a monomial
$y_1^{p_1} \dots y_l^{p_l}$ with $p_k = 1$ then
\begin{equation}
\omega_i + \omega_j = p_1 \omega_1 + \cdots +p_l \omega_l 
+ \sum_{s=1}^l q_s \alpha_s\tag{2.10}
\end{equation}
for some nonnegative integers $q_1, \cdots, q_l$. \par
Let's first consider the root system $A_l$, and assume $\ 1\le i<k<j\le 
l$.\ We multiply (2.10) by $\omega_1$ to obtain
\begin{equation}
{l-i+1 + l-j+1\over l+1} = {\sum p_s (l-s+1) \over l+1} + q_1 .\tag{2.11}
\end{equation}
If $q_1 \geq 1$ then we obtain inequality
$$
l+1 + k - i - j - \sum_{s\neq k} p_s (l-s+1) \geq l+1.
$$
This is impossible since $k-i-j<0$. Hence $q_1=0$ and
$$
l-i+1 - (j-k) = \sum_{s\neq k} p_s (l-s+1).
$$
From the last equation we conclude that $p_s=0$ for $s\le i$.

Next we multiply the equation (2.10) by $\alpha_1, \dots, \alpha_{i-1}$ to 
prove
recursively that also $q_2=\dots = q_i=0$. The last step is to multiply
the same equation by $\alpha_i$. We obtain
$$
1=-q_{i+1}.
$$
This contradicts nonnegativity of $q$'s. So (2.10) is not possible for 
$\ i<k<j$.\par
For the root system $B_l,\ \ k=l-1$,\ \
we assume (2.10) holds true for $\ 1\le i<l-1,\ j=l$.\ We multiply 
(2.10) by $\w_1$ to obtain:
$$
1+\frac12=p_1+\cdots+p_{l-2}+1+\frac12 p_l+q_1,
$$
which leads to $\ p_1=\cdots=p_{l-2}=q_1=0,\ p_l=1$.\ We now multiply
(2.10) by $\w_{i+1}$\ to obtain
$$
i+(\w_l,\w_{i+1})=i+1+(\w_l,\w_{i+1})+q_{i+1},
$$
which is impossible since 
$\ q_{i+1}$\ is a non-negative integer.
So we proved that (2.10) is not valid for $\
1\le i<l-1,\ j=l$.\par
For the root system $D_l,\ \ k=l-2$,\ similar to the case of $B_l$ we 
can see that when $\ 1\le i<l-2$\ and $\ j=l-1$,\ 
or $\  1\le i<l-2$\ and $\ j=l$\
the relation (2.10) can not be valid. When $\ i=l-1,\ j=l$ \ we 
multiply (2.10) by $\w_1$ to obtain
$$
\frac12+\frac12=p_1+\cdots+p_{l-3}+1+\frac12 p_{l-1}+\frac12 p_l+q_1,
$$
which leads to
$\ p_1=\cdots=p_{l-3}=p_{l-1}=p_l=q_1=0$.\ We now multiply (2.10)
by $\ \w_{l-1}$\ to obtain
$$
\frac14 l+\frac14 (l-2)=\frac 12 (l-2)+q_{l-1},
$$
which leads to $\ q_{l-1}=\frac12$,\ this contradicts to the fact that
$q_{l-1}$ is an integer. So under our assumption on $i, j$, \ (2.10)
is not valid.\par
The first statement of the lemma follows from the above arguments
and from the fact that for the root system $C_l$,\ $R_k$ has only
one component. To prove
the second statement of the lemma, we note that in any component of $R_k$, 
the numbers $d_i$ are distinct and ordered monotonically (see Table 2).
Since $\ \eta^{ij}(y)$\ is
a polynomial of degree $\ d_i+d_j-d_k$,\
we have $\ \eta^{ij}(y)=0$\ \ when $\ \ d_i+d_j<d_k$,\ 
and $\ \eta^{ij}=$constant\ \ when $\ \ d_i+d_j=d_k$,\
note that this happens if the labels $i$ and $j$ are dual to each other
in the sense of Lemma 1.6. So we proved the second statement of the 
lemma. The third statement of the lemma follows from (2.5b) and (2.5c).
Lemma is proved.\end{pf}     
\begin{Cor} $\det(\eta^{ij})$ is a constant, modulo
a sign it equals to $\ \prod_{i=1}^l \eta^{ii^*}$.
\end{Cor}
\begin{pf} For the root system $\ A_l-B_l-C_l-D_l$\ the above
statement follows from Lemma 2.3. \par
For the root system $\ E_8,F_4,G_2$
\ we observe that all the numbers $d_i$
are distinct. We can re-label the simple roots in such a way 
that the numbers
$d_i$ are ordered monotonically. Since 
$\ \eta^{ij}(y)$\ is
a polynomial of degree $\ d_i+d_j-d_k$,\ we have
$\ \eta^{ij}=0$\ when $\
d_i+d_k<d_k$\ and $\ \eta^{ij}=$constant \ when $\ d_i+d_j=d_k$.\ The last
equality holds true when \ $j=i^*$.\ We 
conclude 
that the matrix $(\eta^{ij})$ is triangular, and its anti-diagonal elements
are the numbers $\eta^{ii^*}$. So the statement of the 
corollary holds true.\par
For the root system $E_6$, we have
$$ 
d_1=2,\ d_2=3,\ d_3=4,\ d_4=6,\ d_5=4,\ d_6=2,\ d_7=0.
$$ 
We change the labels as follows
$$
4\mapsto 1,\ 3\mapsto 2,\ 5\mapsto 3,\ 2\mapsto 4,\ 6\mapsto 5,
\ 1\mapsto 6,\ 7 \mapsto 7.
$$
Under the new labels the numbers $d_i$ are ordered as follows:
$$
\tilde d_1=6,\ \tilde d_2=4,\ \tilde d_3=4,\ 
\tilde d_4=3,\ \tilde d_5=2,\ \tilde d_6=2,
\ \tilde d_7=0.
$$
 and the matrix $(\eta^{ij})$ becomes $\ (\tilde\eta^{ij})$.\
 We claim that the matrix $(\tilde \eta^{ij})$ is 
triangular, and its anti-diagonal elements consist of the constant
numbers $\ \eta^{ii^*}$.\ To see this, it suffices to show that
$\ \tilde\eta^{36}=\eta^{15}=0$.\ If $\eta^{15}\ne 0$,\ then 
by using a similar argument as we gave in the proof of Lemma 2.3 we have
$$
\w_1+\w_5=p_1\w_1+\cdots+p_6\w_6+\sum_{s=1}^6 q_s\al_s
$$
with $\ p_4=1$\ and $p_i,\ q_j$ are some non-negative integers. This 
is impossible due to
$$
\w_1+\w_5-\w_4=\frac23\alpha_1+\frac13 \alpha_3+\frac23 \alpha_5
+\frac13 \alpha_6
$$
and the fact that $\ (\w_i,\w_j)>0$\ for $\ 1\le i,j\le 6$.\ 
So we proved that the corollary holds true for the root system $E_6$.\par
For the root system $E_7$,\ we have
$$
d_1=4,\ d_2=6,\ d_3=8,\ d_4=12,\ d_5=9,\ d_6=6,\ d_7=3,\ d_8=0.
$$
Similar to the $E_6$ case, to prove the statement of the corollary,
we only need to prove that $\eta^{26}=0$, this easily follows from 
$$
\w_2+\w_6-\omega_4=-\alpha_1-\frac54\al_2-2\al_3-3\al_4-\frac74\al_5
-\frac12 \al_6-\frac14\al_7.
$$
Corollary is proved.
\end{pf}
\begin{Lem} There exists $\bx^0$ such that 
$\ y^0_j(\bx^0)=0$\ for $\ j\ne k$\ and $\ y^0_k(\bx^0)\ne 0$.
\end{Lem}
\begin{pf} We give the required $\bx^0$ explicitly for all cases
of the root systems. 
As in the proof of the main lemma in section 1, we denote
$\ z^0_1=x^0_1,\ z^0_i=x^0_i-x^0_{i-1}$\  and 
$\ \beta^0_1=x^0_l,\ \beta^0_i=x^0_{l-i+1}-x^0_{l-i+2}$\ 
for $\ 2\le i\le l$.\par
For $A_l$, we take
$$
(z^0_1,\dots,z^0_l)=(0,\frac{1}k,\frac{2}k,\dots,\frac{k-1}k,
c,c-\frac{1}{l-k+1},\dots,c-\frac{l-k-1}{l-k+1}),
$$
where $\ c=\frac {l-2k+1}{2(l-k+1)}$.\par
For $B_l$, we take
$$
(z^0_1,\dots,z^0_l)=
(0,\frac1{l-1},\dots,\frac{l-2}{l-1},\frac{3-l}{4}).
$$\par
For $C_l$, we take
$$
(z^0_1,\dots,z^0_l)=(0,\frac {1}{l},\dots,\frac {l-1}{l}).
$$\par
For $D_l$ we take
$$
(z^0_1,\dots,z^0_l)=(0,\frac1{l-2},\dots,\frac{l-3}{l-2},\frac{4-l}{4},0).
$$\par
For $E_6$ we take
$$
(\beta^0_1,\dots,\beta^0_6)=(\frac23,\frac13,0,-\frac23,-\frac1{12},
-\frac14).
$$\par
For $E_7$ we take
$$
(\beta^0_1,\dots,\beta^0_7)=(\frac34,\frac24,\frac14,0,-\frac56,-\frac16,
-\frac13).
$$\par
For $E_8$ we take
$$
(\beta^0_1,\dots,\beta^0_8)=(\frac45,\frac35,\frac25,\frac15,0,
-1,-\frac14,-\frac5{12}).
$$\par
For $F_4$ we take
$$
(x^0_1,x^0_2,x^0_3,x^0_4)=(0,\frac12,\frac23,\frac12).
$$\par
For $G_2$ we take
$$
(x^0_1,x^0_2)=(\frac12,\frac32).
$$
\par
It is now easy to see from the formulae in (1.20),(1.23)--(1.26) that these 
$\ \bx^0=(x^0_1,\dots,x^0_l)$\ satisfy the requirement of the lemma.
Lemma is proved. 
\end{pf}
\begin{Rem} The $\bx^0$\ given in Lemma 2.4 satisfies  
the following relation:
\begin{equation}
\sg_1\sg_2\cdots\sg_{k-1}\sg_{k+1}\cdots\sg_l(\bx^0)=-\frac1{d_k}\w_k
+\sum_{i=1}^k\av_i+\bx^0,\tag{2.12}
\end{equation}
where $\sg$'s are defined in (1.1).
From this relation and (1.17) we obtain
\begin{equation*}
\begin{aligned}
&\zy_j(\sg_1\cdots\sg_{k-1}\sg_{k+1}\cdots\sg_l(\bx^0))
=n_j^{-1}\,\sum\Sb w\in W\\ (w(\w_j)-\w_j,\w_k)=0\endSb e^{2\pi i(w(\w_j),
-\frac1{d_k}\w_k+\sum_{i=1}^k\av_i+\bx^0)}\\
&=n_j^{-1}\,\sum\Sb w\in W\\ (w(\w_j)-\w_j,\w_k)=0\endSb e^{-\frac{2\pi 
i}{d_k} (\w_j, \w_k)+2\pi i (w(\w_j),\bx^0)}
=e^{-\frac{2\pi i d_j}{d_k}} \zy_j(\bx^0).\end{aligned}\tag{2.13}
\end{equation*}
On the other hand from (1.17) we have
\begin{equation*}
\begin{aligned}
&\zy_j(\sg_1\cdots\sg_{k-1}\sg_{k+1}\cdots\sg_l(\bx^0))\\
&=n_j^{-1}\,\sum\Sb w\in W\\ (w(\w_j)-\w_j,\w_k)=0\endSb e^{2\pi i
(\sg_l\cdots\sg_{k+1}\sg_{k-1}\cdots\sg_1w(\w_j),\bx^0)}
=\zy_j(\bx^0).\end{aligned}\tag{2.14}
\end{equation*}
So from (2.13), (2.14) and the fact that $0<\frac {d_j}{d_k}<1$
\ when $\ j\ne k$\ it follows that \newline $\ \zy_j(\bx^0)=0$\ when
$\ j\ne k$.
\end{Rem}
\begin{pf*}{Proof of the main lemma}
By using Corollary 2.2 we only
need to prove that $\ \det(\eta^{ij})$\ does not vanish.\par
Let's denote  
\begin{equation}
\Psi^+=\{\al \in\Phi^+|(\al,\w_k)=0\},\tag{2.15}
\end{equation}
where $\Phi^+$ is the set of all positive roots of $R$.\ Let's
take $x=(\bx,x_{l+1})=(\bx^0-i\tau\w_k, i\tau)$,\ 
where $\bx^0$ is given by Lemma 2.4,
then from
(2.2) we have
\begin{equation*}
\begin{aligned}
&\det(\frac{\pal y^i(x)}{\pal x_j})\\
&=c\ \exp(2\pi i(d_1+\cdots+d_{l})x_{l+1}-\sum_{\al\in \Phi^+}
\pi i(\al,\bx))\prod_{\beta\in\Phi^+}(e^{2\pi i(\beta,\bx)}-1) \\
&=c' \prod_{\beta\in\Psi^+}
(e^{2\pi i(\beta,\bx^0)}-1)\prod_{\beta\in\Phi^+\setminus\Psi^+}
(e^{2\pi i(\beta,\bx^0)+2\pi (\beta,\w_k)\tau}-1).
\end{aligned}\tag{2.16}
\end{equation*}
here
\begin{equation*}
\begin{aligned}
c'&= c\ e^{-\sum_{\al\in\Phi^+}\pi i(\al,\bx^0)-\sum_{\al\in\Phi^+} 
2\pi (\al,\w_k)\tau}\\
&= c\ e^{-\sum_{\al\in\Phi^+}\pi i(\al,\bx^0)-\sum_{\al\in\Phi^+\setminus
\Psi^+}
2\pi (\al,\w_k)\tau},
\end{aligned}
\end{equation*}
and we have used the identity [B,H]
$$
\frac 12\sum_{\al\in\Phi^+}\al=\sum_{i=1}^l\w_i.
$$
Now let's take the limit $\ \tau\to +\infty$,\ 
by using (2.16) we obtain
\begin{equation*}
\begin{aligned}
&\chi :=\lim_{\tau\to +\infty}\det(\frac{\pal y^i(x)}{\pal x_j})\\
&=c\ e^{-\sum_{\al\in\Phi^+}\pi i(\al,\bx^0)}
 \prod_{\beta\in\Psi^+}
(e^{2\pi i(\beta,\bx^0)}-1)\prod_{\beta\in\Phi^+\setminus\Psi^+}
e^{2\pi i(\beta,\bx^0)}.
\end{aligned}
\end{equation*}
From the explicit form of $\bx^0$ given in Lemma 2.4 and the above 
formula we know that $\ \chi\ne 0$.\par
Finally, by using Lemma 1.2, Lemma 2.1, Lemma 2.3--2.4 and Corollary 2.1 
we have $$
\det(\eta^{ij})=\frac1{(\zy_k(\bx^0)^{l+1}}\lim_{\tau\to +\infty}
\det(g^{ij}(x))=c''\lim_{\tau\to +\infty}(\det(\frac{\pal 
y^i(x)}{\pal x_j}))^2 =c''\chi^2\ne 0,
$$
where $c''$ is a non-zero constant.
The main lemma is proved.\end{pf*}
\begin{Cor} The function $\eta^{ij}$ is equal to a 
nonzero constant if and only if $j$ is dual to $i$, i.e., $j=i^*$.
\end{Cor}
\begin{pf} For the root system of type $\ A_l-B_l-C_l-D_l-E_8-F_4-G_2$,\
the corollary follows from the above Main lemma, Lemma 2.3, Corollary 2.2,
Table 2 and the weighted homogeneity of $\eta^{ij}$.\par
For the root system of type $E_6-E_7$, the corollary follows from the
above Main lemma, Corollary 2.2, Table 2, the weighted homogeneity of
$\eta^{ij}$ and the fact that $\ \eta^{15}=\eta^{36}=0$\ for $E_6$ and
$\ \eta^{26}=0$ \ for $E_7$.\ In the proof of Corollary 2.2 we showed
that $\ \eta^{15}=0$\ for $E_6$ and $\ \eta^{26}=0$\ for $E_7$, we can
show in a same way that $\ \eta^{36}=0$\ for $E_6$. Corollary is proved.
\end{pf}
\begin{Cor} The orbit space $\cal M$ carries a 
flat pencil of metrics
$$
g^{ij}(y)\quad \text {and}\quad \eta^{ij}(y)=\frac{\pal g^{ij}(y)}
{\pal y^k}
$$
and the correspondent contravariant Levi-Civita connections
$$
\Gamma^{ij}_m(y)\quad \text {and}\quad \gamma^{ij}_m(y)=
\frac{\pal\Gamma^{ij}_m(y)}{\pal y^k}.
$$
Particularly, the metric $(\eta^{ij}(y))$ is flat.
\end{Cor}
\begin{pf}
This follows from the linearity of $\ g^{ij}(y)$ \ and $\ \Gamma^{ij}_m$
\ in $\ y^k$ \ and from nondegeneracy of $\ \eta^{ij}(y)$ (cf. [D],
Appendix D). Corollary is proved.
\end{pf}  
\par
We recall [D] that this means that the Levi-Civita connection for
a linear combination of the metrics
\begin{equation}
a\ g^{ij}(y)+b \ \eta^{ij}(y)\tag{2.17}
\end{equation}
must have the form
\begin{equation}
a\ \Gamma^{ij}_m(y)+b\ \gamma^{ij}_m(y)\tag{2.18}  
\end{equation}
for arbitrary values of the constants $\ a,b$,\ and the metric (2.17)
must be flat for any $\ a,b$.\par
Note that $\ g^{ij}(y),\ \Gamma^{ij}_m(y)$,\ and also $\ \eta^{ij}(y),\
\gamma^{ij}_m(y)$\ are weighted homogeneous polynomials in $\ y^1,\dots,
y^l, e^{y^{l+1}}$,\ where
$$
\deg y^j=d_j,\quad 1\le j\le l;\quad \deg e^{y^{l+1}}=1.
$$
\begin{Cor} There exist weighted homogeneous
polynomials 
$$
t^{\al}=t^{\al}(y^1,\dots,y^l,e^{y^{l+1}}),\quad \al=1,\dots,l
$$
of the degrees $d_{\al}$ 
such that the metric $\eta^{ij}(y)$ becomes constant in the coordinates
$\ t^1,\dots,t^l,t^{l+1}=y^{l+1}$,\ and the linear part of
$t^\al$ is equal to $y^\al$.
\end{Cor}
\begin{pf}(cf. [D, page 272]) Local existence of the coordinates
$t^{\al}$ follows from vanishing of the curvature of $\eta^{ij}$
(see Corollary 2.4). The flat coordinates $\ t=t(y)$ are to be
found from the system of linear differential equations
\begin{equation*}
\begin{aligned}
&\eta^{is}\frac{\pal \xi_j}{\pal y^s}+\gamma^{is}_j\xi_s=0,\\
&\frac{\pal t}{\pal y^s}=\xi_s.
\end{aligned}\tag{2.19}
\end{equation*} 
From the Main lemma we know that the inverse matrix $\ (\eta_{ij})
=(\eta^{ij})^{-1}$\ is also polynomial in $\ y^1,\dots,y^l, e^{y^{l+1}}$.\
By using the formulae (2.7) we have
$$
\gamma^{i,l+1}_j=\frac{\pal \Gamma_j^{i,l+1}}{\pal y^k}=0,
$$
it follows that 
$$
t^{l+1}=y^{l+1}
$$ 
is one of the solutions of the system (2.19). We choose remaining solutions
\newline $t^{\al}(y^1,\dots,y^l,e^{y^{l+1}})$\ in such a way that
$$
\frac{\pal t^{\al}}{\pal y^j}(0,\dots,0,0)=\delta^{\al}_j,\quad
\al,j=1,\dots,l.
$$
The solutions $\ t^{\al}(y)$ are power series in $\ y^1,\dots,y^l,
e^{y^{l+1}}$.\ The system (2.19) is invariant w.r.t. the 
transformations
$$
y^j\mapsto c^{d_j} y^j,\quad j=1,\dots,l,\quad 
y^{l+1}\mapsto y^{l+1}+\log c
$$
for any non-zero constant $c$.\ So the functions $\ t^{\al}(y),\
1\le \al\le l$\ are weighted homogeneous of the same degrees $d_{\al}>0$.\
Hence the power series $\ t^{\al}(y)$\ must be polynomials.
Corollary is proved.
\end{pf}
\begin{Cor} In the flat  coordinates $\ 
t^1,\dots,t^{l+1}$ we have\newline
\noindent 1)
$$
\eta^{ij}=\frac{\pal g^{ij}(t^1,\dots,t^{l+1})}{\pal t^k}.
$$
\noindent 2)\ \ $\eta^{ij}$ is equal to a nonzero constant if and only if
$j=i^*$, and
$$ 
\eta^{ii^*}(t^1,\dots,t^{l+1})=\eta^{ii^*}(y^1,\dots,y^{l+1}).
$$
\end{Cor}
\begin{pf} From Corollary 2.5 we have
$$
\frac{\pal}{\pal t^k}=\frac{\pal}{\pal y^k},
$$
which leads to the first statement of the corollary. \par
The second statement of the corollary follows from the fact that
the linear part of $t^\alpha$ is $y^\alpha$. \ Corollary is proved
\end{pf} 
\par
It follows from our normalization of the flat coordinates that
$$
\eta^{l+1,\alpha}=\delta^\alpha_k.
$$
\begin{Rem} For the orbit spaces of finite reflection
groups flat coordinates were constructed by Saito, Yano and Sekiguchi
in [SYS] (see also [S]).\end{Rem}
\par
We recall now the definition of Frobenius manifold.\par
\noindent {\bf {Definition.}} (Smooth, polynomial etc.) 
{\it {Frobenius
structure}} on a $n$-dimensional manifold $M$ consists of:\newline
1)\ \ a structure of commutative Frobenius algebra with a unity
$e$ on the tangent plane $\text{T}_{\text {t}} M$ that depends 
smoothly, polynomially etc. on $t\in M$.\ (We recall that a 
commutative associative algebra $A$ is called Frobenius algebra
if it is equipped with a symmetric nondegenerated bilinear
form $\ <\ ,\ >$\ satisfying the invariance condition
$$
<a\ b, c>=<a, b\ c>
$$
for any $\ a, b, c\in A$.)\newline
2)\ \ A vector field $E$ is fixed on $M$.\ We will call it {\it {Euler 
vector field}}.\newline
These objects must satisfy the following properties:\newline
i)\ \ The curvature of the invariant metric $\ <\ ,\ >$\ on $M$\
 is equal to zero;\newline
ii)\ \ denoting $\bn$ the Levi-Civita connection for $\ <\ ,\ >$,\
we require that 
\begin{equation}
\bn\ e=0;\tag{2.20}
\end{equation}
\noindent 
iii)\ \ the four-tensor
$$
d(a_1,\dots,a_4) :=(\bn_{a_4} c)(a_1,a_2,a_3),
$$
where
$$
c(a_1,a_2,a_3)=<a_1a_2,a_3>,
$$
must be symmetric w.r.t. any permutation of the vectors $\ a_1,\dots,
a_4$.\newline
iv)\ \ The vector field $E$ must be linear w.r.t. $\bn$:
\begin{equation}
\bn\bn E=0.\tag{2.21}
\end{equation}
The eigenvalues $\ q_1,\dots,q_n$\ of the linear operator $\ 
\text {id}-\bn E$\ are called {\it {charges}} of the Frobenius
structure.\newline
v)\ \ The Lie derivative $\ \cal L_{E}$\  along the vector field 
$E$ must act by rescalings
\begin{equation*}
\begin{aligned}
&\cal L_{E} e=-e,\\
&\cal L_{E} (a\ b)-(\cal L_{E} a)b-a(\cal L_{E} b)=ab,\\
&\cal L_{E}<a,b>-<\cal L_{E} a,b>-<a,\cal L_{E} b>=(2-d)<a,b>,
\end{aligned}\tag{2.22}
\end{equation*}
for arbitrary vector fields $\ a,\ b$\ and for
certain constant $d$.\ (Observe that, due to (2.20) and (2.22),\
zero and $d$ are among the charges $q_1,\dots,q_n).$
\par
A manifold $M$ with a Frobenius structure on it is called a 
{\it { Frobenius manifold}}.\par
If we choose locally flat coordinates $\ t^1,\dots,t^n$\
for the  invariant metric, then the condition iii) provides local
existence of a function $\ F(t^1,\dots,t^n)$\ such that
\begin{equation}
<a\ b,c>=a^{\al}b^{\beta}c^{\gamma}\frac{\pal^3 F}{\pal t^{\al}
\pal t^{\beta}\pal t^{\gamma}}\tag{2.23}
\end{equation}
for any three vectors $\ a=a^{\al}\frac {\pal}{\pal t^{\al}},\
b=b^{\beta}\frac {\pal}{\pal t^{\beta}},\ c=c^{\gamma}\frac {\pal}{\pal 
t^{\gamma}}$.\
Choosing the coordinate $\ t^1$\ along the unity vector field $\ e$\
we obtain
\begin{equation}
{\partial^3 F(t^1, \dots, t^n) \over \partial t^1 \partial t^\alpha \partial t^\beta}
=\eta_{\alpha\beta}\tag{2.24a}
\end{equation}
for a constant symmetric nondegenerate matrix $\ \left( 
\eta_{\alpha\beta}\right)$\ coinciding with the metric $\ <\ , \ >$\
in the chosen coordinates.
Associativity of the algebras implies an overdetermined system of 
equations for the function $F$
\begin{equation}
{\partial^3 F\over \partial t^\alpha \partial t^\beta \partial t^\lambda} 
\eta^{\lambda\mu} {\partial^3 F 
\over \partial t^\mu  \partial t^\gamma \partial t^\delta} =
{\partial^3 F \over \partial t^\gamma \partial t^\beta \partial t^\lambda}
\eta^{\lambda\mu} {\partial^3 F 
\over \partial t^\mu \partial t^\alpha \partial t^\delta} \tag{2.24b}
\end{equation}
for arbitrary $\ \alpha, \ \beta, \ \gamma, \ \delta$\ from $1$ to 
$n$. 
The components of the Euler vector field
$E$ in the basis $\ \frac{\pal}{\pal t^{\al}}$\ are linear 
functions of $\ t^1,\dots,t^n$. They enter into the following scaling 
condition for the function $F$
\begin{equation}
{\cal L}_E F = (3-d) F + \ \ \text {quadratic polynomial in} \ \ t
\tag{2.24c}
\end{equation}
The system (2.24a-c) is just the {\it {WDVV equations of 
associativity}} being equivalent to our definition of Frobenius
manifold in the chosen system of local coordinates. 
\par
We recall also an important construction of {\it { intersection
form}} of a Frobenius manifold. This is a symmetric bilinear form
$\ (\ ,\ )^*$\ on $\ \text {T}^*M$\ defined by the formula
\begin{equation*}
(w_1,w_2)^*=i_E (w_1\cdot w_2),\notag
\end{equation*}
here the product of two $1$-forms $\ w_1, w_2$\ at a point $\ t\in M$\
is defined by using the algebra structure on $\text {T}_{\text {t}}M$
and the isomorphism
$$
\text {T}_{\text {t}}M\to\text {T}_{\text {t}}^*M
$$
established by the invariant metric $\ <\ ,\ >$.\
Choosing the flat coordinates 
$\ t^1,\dots,t^n$\ for the invariant metric, we can rewrite the 
definition of the intersection form as
\begin{equation}
(dt^{\al},dt^{\beta})^*=\cal L_E F^{\al\beta},\tag{2.25}
\end{equation}
where
$$
F^{\al\beta}=\eta^{\al\al'}\eta^{\beta\beta'}\frac{\pal^2 F(t)}
{\pal t^{\al'}\pal t^{\beta'}},
$$
and the function $ F(t)$ is defined in (2.23). According to the 
general theory of Frobenius manifolds, $\ (\ , \ )^*$\
defines a new flat metric on the open subset of $M$ where $\ \det(\ 
,\ )^*\ne 0$.\ The {\it {discriminant}} $\ \Sigma = \{ t \ |
\ \det(\ , \ )^*_t =0\}$\ is a proper analytic subvariety in an 
analytic Frobenius manifold $M$. The holonomy of the local Euclidean 
structure defined on $M\setminus \Sigma$ by the intersection form
$\ (\ , \ )^*$ gives a representation
$$
\pi_1 \left( M\setminus \Sigma \right)\ \  \to \ \ \text {Isometries} 
\ V 
$$
where $V$ is the standard complex Euclidean space. The image of this 
representation is called {\it {monodromy group}} of the Frobenius 
manifold.
\begin{Thm} There exists a unique Frobenius 
structure on the orbit space $\ \cal M=\cal M(R,k)$\ polynomial
in $\ t^1,\dots,t^l, e^{t^{l+1}}$\ such that \newline
1)\ \ the unity vector field coincides with $\ \frac {\pal}{\pal y^k}
=\frac {\pal}{\pal t^k}$;\newline
2)\ \ the Euler vector field has the form 
\begin{equation}
E=\frac 1{2\pi i d_k}\frac {\pal}{\pal x_{l+1}}
=\sum_{\al=1}^l\frac{d_{\al}}{d_k}\ t^{\al}\frac{\pal}{\pal t^{\al}}
+\frac1{d_k}\frac{\pal}{\pal t^{l+1}}.\tag{2.26}
\end{equation}
\newline
3)\ \ The intersection form of the Frobenius structure coincides with
the metric $\ (\ ,\ )^{\sptilde}$\ on $\ \cal M$.\par
\end{Thm}
Observe that the charges $\ q_1,\dots,q_l$\ are
$$
q_j=\frac{(\w_k-\w_j,\w_k)}{(\w_k,\w_k)},\quad j=1,\dots,l,
\quad q_{l+1}=d=1.
$$
\begin{Cor} The monodromy group of ${\cal M}(R,k)$ 
is isomorphic to the group \newline $\widetilde W^{(k)}(R)$.
\end{Cor}
The proof of the theorem will be based on the following lemmas (cf.
[D, pp. 273-275]):
\begin{Lem} In the coordinates $\ t^1,\dots,t^{l+1}$
\begin{equation*}
\begin{aligned}
&g^{\al ,l+1}=\frac{d_{\al}}{d_k} t^{\al},\quad \al=1,\dots,l,
\quad g^{l+1,l+1}=\frac1{d_k},\\
&\Gamma^{l+1, \al}_{\beta}=\frac{d_{\al}}{d_k}\delta_{\beta}^{\al},
\quad 1\le \al,\beta\le l+1.
\end{aligned}
\end{equation*}
\end{Lem}
The proof of this lemma is straightforward using (2.5b),(2.5c),(2.7)
,$\ t^{l+1}=y^{l+1}$\ and the quasi-homogeneity of $\ t^1,\dots,t^l$.
\begin{Lem} There exists a unique weighted homogeneous
polynomial\newline
$\ G=G(t^1,\dots,t^{k-1},t^{k+1},\dots,t^l,e^{t^{l+1}})$\ of the degree 
$2d_k$ such that the function
$$
F=\frac12 (t^k)^2t^{l+1}+\frac12 t^k\sum_{\al,\beta\ne k}\eta_{\al\beta}
t^{\al}t^{\beta}+G
$$
satisfies the equations
\begin{equation}
(dt^{\al},dt^{\beta})^{\sptilde}=\cal L_E F^{\al\beta}.\tag{2.27}
\end{equation}
\end{Lem}
\begin{pf} Let $\ \Gamma^{\alpha\beta}_{\gamma}(t)$\ be the coefficients
of the Levi-Civita connection for the metric $\ (\ ,\ )^{\sptilde}$\
in the coordinates $\ t^1,\dots,t^{l+1}$.\ We use now the theory of
flat pencils of metrics (see [D, Appendix D]). According to Proposition
D.1 of [D] we can represent these functions as
\begin{equation}
\Gamma^{\alpha\beta}_{\gamma}(t)=\eta^{\alpha\ve}\pal_{\ve}\pal_{\gamma}
f^{\beta}(t)\tag{2.28}
\end{equation}
for some functions $\ f^{\beta}(t)$.\ From the weighted homogeneity of 
$\ \Gamma^{\alpha\beta}_{\gamma}(t)$\ and Corollary 2.6 we 
obtain that
$$
\partial_\alpha \partial_\gamma \left( {\cal L}_E f^\beta - {d_k + 
d_\beta \over d_k} f^\beta \right) =0
$$
for any $\alpha, ~\beta$. So
\begin{equation}
{\cal L}_E f^\beta (t) = {(d_\beta +d_k)\over d_k}f^\beta(t)
+A^\beta_\sigma t^\sigma +B^\beta\tag{2.29}
\end{equation}
for some constants $A^\beta_\sigma$, $B^\beta$. 
Doing a transformation
$$f^\beta(t)\mapsto \tilde f^\beta (t)=f^\beta(t) +R^\beta_\lambda 
t^\lambda+ Q^\beta
$$
we can kill all the coefficients $A^\beta_\sigma$, $B^\beta$ in (2.29)
except $A^{l+1}_k$. Indeed, after the transformation we obtain
\begin{equation*}
\begin{aligned}
&{\cal L}_E \tilde f^\beta (t) = {(d_\beta+d_k)\over d_k}
\tilde f^\beta (t)+
\sum_{\gamma =1}^l \left[ R^\beta_\gamma {d_\gamma -d_k 
-d_\beta \over d_k}+A^\beta_\gamma \right]t^\gamma \\
&+ {1\over d_k}R^\beta_{l+1}+ B^\beta -{d_k +d_\beta \over d_k}Q^\beta
+\left[ A_{l+1}^\beta - {d_k + d_\beta \over d_k} 
R^\beta _{l+1}\right] t^{l+1}.
\end{aligned}
\end{equation*}
The function $\tilde f^\beta (t)$ does still satisfy (2.28). Choosing
\begin{equation*}
\begin{aligned}
&R^\beta_{l+1} ={d_k \over d_k +d_\beta} A^\beta_{l+1},
\\ &
Q^\beta ={d_k \over d_k +d_\beta} \left[ B^\beta + {1 \over d_k} 
R^\beta_{l+1}\right]\end{aligned}
\end{equation*}
we kill the constant term in the r.h.s. of (2.29) and the term linear in 
$t^{l+1}$.\par

To kill other linear terms we are to put
$$R^\beta_\gamma = {d_k \over d_\beta +d_{\gamma^*}}A^\beta_\gamma
$$
where $\gamma^*$ is the index dual to $\gamma$ in the sense of 
duality (1.27). We can do this unless $d_\beta = d_{\gamma^* }=0$.
The last equation holds only for $\beta = \gamma^* = l+1$, i.e. for
$\beta=l+1$, $\gamma=k$. So, we can kill all the linear terms but 
$\ A^{l+1}_k$\ in (2.29).\par

Thus for $\beta \neq l+1$ the polynomials $f^\beta (t)$ can be 
assumed to be homogeneous of the degree $d_\beta + d_k$.\par
We show now that for $1\leq \beta \leq l$ the functions $f^\beta$
are polynomials in $t^1, \dots, t^l$,\newline
$\exp t^{l+1}$. We already know 
that this is true for the Christoffel coefficients 
$\Gamma_\gamma^{\alpha\beta}$. Let us denote
$$
\eta_{\alpha\epsilon} \Gamma^{\epsilon\beta}_{l+1} (t) = \sum_{m=0}^N
C_{\alpha, m}^\beta \exp m t^{l+1} 
\equiv \partial_\alpha \partial_{l+1} f^\beta (t)
$$
where the coefficients $C_{\alpha, m}^\beta$ are polynomials in $t^1, 
\dots, t^l$ and $N$ is certain positive integer. From compatibility
$$
\partial_{l+1}\left( \partial _\alpha \partial_{l+1} f^\beta \right)
= \partial_\alpha \left( \partial_{l+1}^2 f^\beta \right)
$$
we obtain that 
$$
\partial_\alpha C_{l+1, 0}^\beta =0, ~\alpha = 1, \dots, l.
$$
So $C_{l+1, 0}^\beta$ is a constant. But $\partial_{l+1}^2 f^\beta$ 
must be
a weighted homogeneous polynomial of the positive degree $d_k + 
d_\beta$. Hence $C_{l+1, 0}^\beta =0$ and we obtain
$$
f^\beta = \sum_{m=1}^N {1 \over m^2} C_{l+1, m}^\beta \exp m t^{l+1}
+ t^{l+1} D^\beta + H^\beta
$$
for some new polynomials $D^\beta = D^\beta (t^1, \dots, t^l)$ and 
$H^\beta = H^\beta (t^1, \dots, t^l)$. Since the derivatives 
$\partial_\alpha \partial_\gamma f^\beta$ must not contain terms 
linear in $t^{l+1}$, the polynomial $D^\beta$ is at most linear
in $t^1, \dots, t^l$. Using homogeneity of $f^\beta$ we conclude
that $D^\beta =0$.

The coefficient $\ \Gamma^{\al\beta}_{\gamma}(t)$\ must also
satisfy the conditions [D]
\begin{equation}
g^{\al\sigma}\Gamma^{\beta\gamma}_{\sg}=g^{\beta\sg}\Gamma^{\al
\gamma}_{\sg}.\tag{2.30}
\end{equation}
For $\ \al=l+1$\ because of (2.28),(2.30) and Lemma 2.5  we obtain
$$
\cal L_E(\eta^{\beta\ve}\pal_{\ve}f^{\gamma})=\frac{d_{\gamma}}{d_k}
g^{\beta\gamma}.
$$
Because of $\ \deg f^{\gamma}=d_{\gamma}+d_k$\ we have
$\ \deg(\eta^{\beta\ve}\pal_{\ve} f^{\gamma})=d_{\beta}+d_{\gamma}$\
for $\ \gamma\ne l+1$,\
so
\begin{equation}
(d_{\gamma}+d_{\beta})\eta^{\beta\ve}\pal_{\ve}
f^{\gamma}=d_{\gamma} g^{\beta\gamma},\quad \gamma\ne l+1.\tag{2.31}
\end{equation}
We introduce functions $\ F^{\gamma}$ \ for $\ \gamma\ne l+1$\ putting
$$
F^{\gamma}=\frac{d_k}{d_{\gamma}} f^{\gamma}.
$$
From (2.31) we obtain the equation
\begin{equation}
\eta^{\beta\ve}\pal_{\ve} F^{\gamma}=\eta^{\gamma\ve}\pal_{\ve}F^{\beta},
\quad 1\le \gamma,\beta\le l.\tag{2.32}
\end{equation}
From (2.32) it follows that a function $F(t)$ exists such that
\begin{equation}
F^\gamma = \eta^{\gamma\epsilon} \partial_\epsilon F, 
\quad 1\leq\gamma\leq l.\tag{2.33}
\end{equation}
The dependence of $F$ on $t^k$ is not determined from (2.33). However, 
 putting $\beta = l+1$ in (2.31) we obtain
\begin{equation}
\partial_k F^\gamma = t^\gamma , ~~1\leq \gamma \leq l,\tag{2.34}
\end{equation}
from (2.33), (2.34) and Corollary 2.6 we obtain
\begin{equation*}
\begin{aligned}
&\partial_{l+1} \left( \partial_k F\right) = t^k \\
&\partial_\gamma \left( \partial_k F \right) = 
\sum_{\al=1}^l \eta_{\gamma\al} t^{\al},\quad \gamma\neq k, l+1.
\end{aligned}
\end{equation*}
Hence we have
$$\partial_k F = t^k t^{l+1} + {1\over 2} \sum_{\alpha, \beta \neq k,l+1}
\eta_{\alpha\beta} t^\alpha t^\beta + g(t^k)
$$
for some function $g$. Shifting $F\mapsto F+ \int g(t^k) \, dt^k$
we can kill this function, and the equations in (2.33) still hold 
true due to $\ \eta^{ik}=\delta_{i,l+1}$.\
We obtain the representation
\begin{equation}
F = {1\over 2} \left(t^k\right)^2 t^{l+1} +{1\over 2} t^k 
\sum_{\alpha, \beta \neq k,l+1} \eta_{\alpha\beta} t^\alpha t^\beta + G
\tag{2.35}
\end{equation}
with some $G$ independent on $t^k$.\par
From the definition (2.33) of $F$ and the weighted 
homogeneity of $f^\gamma,\
\gamma\ne l+1$ we obtain that
$$
{\cal L}_E F(t) = 2 F(t) + a(t^k)
$$
for some unknown function $a$. Using the duality condition (1.27) 
and Corollary 2.6 we 
obtain
$$
{\cal L}_E F(t)= {1\over 2 d_k} \left( t^k\right)^2 + \left( t^k 
\right)^2 t^{l+1} + t^k \sum_{\alpha, \beta \neq k,l+1} 
\eta_{\alpha\beta} t^\alpha t^\beta +{\cal L}_E G(t), 
$$
or, equivalently
$$
{\cal L}_E G(t) = 2 G(t) + a(t^k) - {1\over 2 d_k} (t^k)^2.
$$
But ${\cal L}_E G(t)$ does not depend on $t^k$. We obtain 
$$
a(t^k) = {1\over 2 d_k} (t^k)^2 + c
$$
for some constant $c$. Killing the constant by a shift, we obtain 
that\newline $G\left( t^1, \dots, \hat t^k,\dots, t^{l+1}\right)$ is a 
weighted homogeneous function of the degree $2 d_k$. The above 
conditions determine this function uniquelly. Clearly $G$ is
a polynomial in $t^1, \dots, \hat t^k, \dots, t^l, \exp t^{l+1}$
(it was obtained by integrating polynomials).\par
Substituting $F(t)$ into (2.31) we obtain (2.27) for 
$1\leq \alpha \leq l$. Finally, for $\alpha = \beta = l+1$ the 
equation (2.27) reads
$${1\over d_k}= {\cal L}_E {\partial^2 F\over \partial t^k 
\partial t^k}.
$$
This follows immediately from the explicit form of $F$. Lemma is 
proved.
\end{pf}
\begin{Lem} The functions 
$$
c^{\al\beta}_{\gamma}(t)=\eta^{\al\al'}\eta^{\beta\beta'}
\frac{\pal^3 F(t)}{\pal t^{\al'}\pal t^{\beta'}\pal t^{\gamma}}
$$
are weighted homogeneous polynomials in $\ t^1,\dots,t^l, e^{t^{l+1}}$\
of the degreees $\ d_{\al}+d_{\beta}-d_{\gamma}$.\ They satisfy the 
associativity equations
$$
c^{\al\beta}_{\sg}c^{\sg\gamma}_{\delta}=c^{\al\sg}_{\delta}
c^{\beta\gamma}_{\sg}.
$$
\end{Lem}
\begin{pf} Weighted homogeneity of $\ c^{\al\beta}_{\gamma}$\
for $\ \al$\ or $\ \beta\ne l+1$\ follows from those of
the functions $\ \Gamma^{\al\beta}_{\gamma}$\ since
\begin{equation}
\Gamma^{\al\beta}_{\gamma}=\frac{d_{\beta}}{d_k}c^{\al\beta}_{\gamma},
\tag{2.36}
\end{equation}
which follows from (2.28). Due to (2.35)
 we also have
\begin{equation}
c^{\al,l+1}_{\gamma}=\delta^{\al}_{\gamma}.\tag{2.37}
\end{equation}
This is also a weighted homogeneous function.\par
To prove associativity we use again the theory of linear pencils of
the flat metrics $\ g^{\al\beta},\ \eta^{\al\beta}$.\ \ Using
[D, eq. (D.2)] we obtain
\begin{equation}
\Gamma^{\al\beta}_{\sg}\Gamma^{\sg\gamma}_{\delta}=\Gamma^{\al\gamma}_{\sg}
\Gamma^{\sg\beta}_{\delta}.\tag{2.38}
\end{equation}
Substitute (2.36),(2.37) into (2.38) we obtain 
$$
c_\sigma^{\alpha\beta}c_\delta^{\sigma\gamma}=c_\sg^{\alpha\gamma}
c_\delta^{\sg\beta}.
$$
Due to commutativity of the multiplication we obtain
needed associativity.
Lemma is proved.
\end{pf}
\begin{pf*}{Proof of Theorem 2.1} From Lemma 2.5-2.7 we know that 
we only need to verify that $\ e=\frac {\pal}{\pal t^k}$\
is the unity of the algebra and that
$$
\cal L_E e=-e.
$$
This is very simple. Theorem is proved.
\end{pf*}
\begin{Rem}
Any orthogonal map $T: V \to V$ preserving the set of simple roots 
defines an isomorphism of Frobenius manifolds
$$
{\cal M}(R,k) \to {\cal M}(R,k')
$$
where 
$$
T(\alpha_k) = \alpha_{k'}.
$$
Particularly, for the root system of the type $A_l$ we obtain an 
isomorphism
$$
{\cal M}(A_l, k) \simeq {\cal M}(A_l, l-k+1)
$$
corresponding to the reflection of the Dynkin graph w.r.t. the 
center.  
\end{Rem}

\vskip 0.4 cm
\par 
We now give some examples to illustrate our above construction. For
convenience, instead of $\ t^1,\dots,t^{l+1}$\ we will denote the flat 
coordinates of the metric $(\eta^{ij})$ by $\ t_1,\dots,t_{l+1}$,
and we will also denote $\ \pal_i=\frac {\pal}{\pal t_i}$\ in these examples.
\par
\vskip 0.4cm
\noindent{\bf {Example 2.1.}}\ \
For the root system of the type $A_1$ the affine Weyl group acts on 
$x_1$-line by transformations
$$
x_1 \mapsto \pm x_1 + m
$$
for an integer $m$.
Our extension $\tilde W(A_1)$ consists of transformations of 
$(x_1,x_2)$-plane of the form
$$
(x_1,x_2)\mapsto \left( \pm x_1 + m + \frac12 n, x_2-n \right)
$$
for arbitrary integers $m$, $n$.
Basic invariants of this group bounded along the lines
$$
(x_1,x_2) = (x_1^0 -\frac12 i \tau, x_2^0 + i \tau ), \ \ \tau \to 
+\infty
$$
are
$$
t_1 =2\, e^{\pi i x_2} \cos 2 \pi x_1  \ \ \text {and} \ \  
e^{2 \pi i x_2}.
$$
The extended invariant metric on the dual space has the matrix
$$
((dx_i,dx_j)^{\sptilde})={1\over 8 \pi^2}  \pmatrix 
1 & 0 \\
0 & -4 
\endpmatrix.
$$
In the coordinates $t_1$, $t_2= 2\pi i x_2$ this metric has the matrix
$$
\left( g^{\alpha\beta}\right) =  \pmatrix
2\,e^{t_2} & t_1 \\
t_1 & 2  
\endpmatrix .
$$
So the Frobenius structure is determined by the function
$$
F = \frac12 t_1^2 t_2 + e^{t_2}.
$$
Up to normalization this is the free energy of {\bf CP}$^1$ 
topological sigma-model [D].
\par
\vskip 0.5cm
\noindent{\bf {Example 2.2.}}\ \
 Let $R$ be the root system 
$A_2$, we take $\ k=1$,\ then $\ d_1=\frac 23,\ \ d_2=\frac 13$,\ and
\begin{equation*}
\begin{aligned} 
&y^1=e^{\frac 43\pi i x_3} (e^{2\pi i x_1}+e^{-2\pi i x_2}+
     e^{2\pi i(x_2-x_1)});\\
&y^2=e^{\frac23\pi i x_3} (e^{2\pi i x_2}+e^{-2\pi i x_1}+
     e^{2\pi i(x_1-x_2)});\\
&y^3=2\pi i x_3.
\end{aligned}
\end{equation*}
The metric $\ (\ ,\ )^{\sptilde}$\ has the form
$$
((dx_i,dx_j)^{\sptilde})=\frac1{12\pi^2}\pmatrix 
2&1&0\\ 1& 2&0\\ 0&0&-\frac 92
\endpmatrix.
$$ 
The flat coordinates 
$\ 
t_1=y^1,\ \  t_2=y^2,\ \ t_3=y^3
$,\
the intersection form is given by
$$
(g^{ij})=\pmatrix 2t_2 e^{t_3}& 3e^{t_3} & t_1\\ 
3e^{t_3} & 2 t_1-\frac12 t_2^2 & \frac12 t_2\\ 
t_1 &\frac12 t_2&\frac32\endpmatrix.
$$
The free energy
$$
F=\frac12 t_1^2 t_3+\frac14 t_1 t_2^2+t_2 e^{t_3}-\frac1{96} t_2^4,
$$
and the Euler vector field reads
$$
E=t_1\pal_1+\frac12 t_2\pal_2+\frac32\pal_3.
$$ \par
\vskip 0.5cm
\noindent {\bf {Example 2.3.}}\ \ Let $R$ be the root system $C_2$,
then $\ k=2,\ \ d_1=1,\ d_2=2$,\ and
\begin{equation*}
\begin{aligned}
&y^1=e^{2\pi i x_3}(e^{2\pi i x_1}+e^{-2\pi ix_1}+e^{2\pi i (x_2-x_1)}+
    e^{-2\pi i(x_2-x_1)}).\\
&y^2=e^{4\pi i x_3}(e^{2\pi i x_2}+e^{-2\pi i x_2}+e^{2\pi i (2x_1-x_2)}+
     e^{-2\pi i (2x_1-x_2)}),\\
&y^3=2\pi i x_3.
\end{aligned}
\end{equation*}
The metric $\ (\ ,\ )^{\sptilde}$\ has the form
$$
((dx_i,dx_j)^{\sptilde})=\frac1{4\pi^2}\pmatrix
1&1&0\\ 1&2&0\\ 0&0&-\frac12
\endpmatrix.
$$
The flat coordinates
$\
t_1=y^1,\ \  t_2=y^2+2 e^{2 y^3},\ \ t_3=y^3
$,\
the intersection form is given by
$$
(g^{ij})=\pmatrix 2t_2-\frac12 t_1^2+4 e^{2 t_3}&
6t_1 e^{2t_3}&\frac12 t_1\\ \\
6t_1 e^{2t_3}&8 e^{4t_3}+4t_1^2 e^{2 t_3}&t_2\\ 
\frac12 t_1& t_2&\frac12\endpmatrix.
$$ 
The free energy
$$
F=\frac14 t_1^2 t_2+\frac12 t_2^2 t_3-\frac1{96} t_1^4
+\frac12 t_1^2 e^{2t_3}+\frac14 e^{4t_3},
$$
and the Euler vector field reads
$$
E=\frac12 t_1\pal_1+ t_2\pal_2+\frac12\pal_3.
$$
\vskip 0.4cm
\begin{Rem}  The root system $B_2$ gives a Frobenius 
structure
which is isomorphic to the one given by the root system $C_2$.
\end{Rem}
\par
\vskip 0.5 cm
\noindent {\bf {Example 2.4.}}\ \ Let $R$ be the root system $G_2$,
then $\ k=2,\ \ d_1=3,\ d_2=6$,\ and
\begin{equation*}
\begin{aligned}
y^1=&e^{6\pi i x_3}(e^{2\pi i x_1}+e^{-2\pi ix_1}+e^{2\pi i (2x_1-x_2)}+
    e^{-2\pi i(2x_1-x_2)}+\\
    &+e^{2\pi i (x_1-x_2)}+e^{-2\pi i (x_1-x_2)}).\\
y^2=&e^{12\pi i x_3}(e^{2\pi i x_2}+e^{-2\pi i x_2}+e^{2\pi i (3x_1-2x_2)}+
     e^{-2\pi i (3x_1-2x_2)}+\\
    &+e^{2\pi i (x_2-3x_1)}+
      e^{-2\pi i (x_2-3x_1)}),\\
y^3=&2\pi i x_3.
\end{aligned}
\end{equation*}
The metric $\ (\ ,\ )^{\sptilde}$\ has the form
$$
((dx_i,dx_j)^{\sptilde})=\frac1{4\pi^2}\pmatrix
2&3&0\\ 3&6&0\\ 0&0&-\frac16
\endpmatrix.
$$
The flat coordinates
$\
t_1=y^1+2 e^{3 y^3},\ \  t_2=y^2+3 y^1 e^{3 y^3}+6 e^{6y^3},\ \ t_3=y^3
$,\
the intersection form is given by
$$
(g^{ij})=\pmatrix 2t_2-\frac12 t_1^2+8t_1 e^{3t_3}+4 e^{6t_3}&
9(2t_1 e^{6t_3}+t_1^2 e^{3t_3})&\frac12 t_1 \\
9(2t_1 e^{6t_3}+t_1^2 e^{3t_3})&
6(6 t_1^2 e^{6t_3}+t_1^3 e^{3t_3}+4e^{12t_3})&t_2\\ 
\frac12 t_1&t_2&\frac16\endpmatrix.
$$  
The free energy
$$
F=\frac14 t_1^2 t_2+\frac12 t_2^2 t_3-\frac1{96} t_1^4+
\frac13 t_1^3 e^{3t_3}+\frac12 t_1^2 e^{6t_3}+\frac1{12} e^{12 t_3},
$$
and the Euler vector field
$$
E=\frac12 t_1\pal_1+t_2\pal_2+\frac16 \pal_3.
$$
\vskip 0.6cm
\begin{Rem}
All the above examples was found in [D] (although the relation of 
Example 2.4 to an extension of the affine Weyl group of the type 
$G_2$ was not proved in [D]). It is important to notice that these 
are all $n$-dimensional Frobenius manifolds with $n\leq 3$ being 
polynomial in $t_1, \dots, t_{n-1}, \exp t_n$ with $\deg \exp t_n >0$
(see [D], Appendix A). It would be natural to conjecture that our 
construction gives all such Frobenius manifolds (with semisimplicity 
condition [D] added) for any $n$.
\end{Rem}

We proceed now to the list of all four-dimensional Frobenius 
manifolds given by our construction.

\vskip 0.5cm
\noindent {\bf {Example 2.5.}}\ \ Let $R$ be the root system $A_3$,
take $\ k=1$,\ then $\ d_1=\frac34,\ d_2=\frac12,\ d_3=\frac14$,\ and
\begin{equation*}
\begin{aligned}
y^1=&e^{\frac32\pi i x_4}(e^{2\pi i x_1}+e^{2\pi i(x_2-x_1)}
+e^{2\pi i(x_3-x_2)}+e^{-2\pi ix_3});\\
y^2=&e^{\pi i x_4}(e^{2\pi i x_2}+e^{-2\pi i x_2} 
     +e^{2\pi i (x_1-x_3)}+e^{-2\pi i(x_1-x_3)}+\\
     &+e^{2\pi i (x_1+x_3-x_2)}+e^{-2\pi i(x_1+x_3-x_2)} );\\
y^3=&e^{\frac12 \pi i x_4}(e^{2\pi i x_3}+e^{2\pi i(x_1-x_2)}
+e^{2\pi i(x_2-x_3)}+e^{-2\pi ix_1});\\
y^4=&2\pi i x_4.
\end{aligned}
\end{equation*}
The metric $\ (\ ,\ )^{\sptilde}$\ has the form
$$
((dx_i,dx_j)^{\sptilde})=\frac1{16\pi^2}\pmatrix
3&2&1&0\\ 2&4&2&0\\ 
1&2&3&0\\
0&0&0&-\frac{16}3
\endpmatrix.
$$
The flat coordinates
$\
t_1=y^1,\ \  t_2=y^2-\frac16 (y^3)^2,\ \ t_3=y^3,\ \ t_4=y^4
$,\
the intersection form is given by
$$
(g^{ij})=\pmatrix 2(t_2+\frac16 t_3^2) e^{t_4}
&\frac53 t_3e^{t_4}& 4e^{t_4}&t_1\\ 
\frac53 t_3e^{t_4}&\frac29 t_2t_3^2-\frac23 t_2^2-\frac1{54} t_3^4
+4e^{t_4}& \frac1{18}t_3^3-t_2 t_3+3t_1&\frac23 t_2\\ 
4e^{t_4}&\frac1{18}t_3^3-t_2 t_3+3t_1&2t_2-\frac13 t_3^2&\frac13 t_3\\ 
t_1&\frac23 t_2&\frac13 t_3&\frac43
\endpmatrix.
$$
The free energy
$$
F=\frac12 t_1^2 t_4+\frac13 t_1 t_2 t_3+\frac1{18} t_2^3-\frac1{36} t_2^2 
t_3^2+\frac1{648} t_2 t_3^4
-\frac1{19440} t_3^6
+(t_2+\frac16 t_3^2) e^{t_4},
$$
and the Euler vector field reads
$$
E=t_1\pal_1+\frac23 t_2\pal_2+\frac13 t_3\pal_3+\frac43\pal_4.
$$
\vskip 0.5cm
\noindent {\bf {Example 2.6.}}\ \ Let $R$ be the root system $A_3$,
take $\ k=2$,\ then $\ d_1=\frac12,\ d_2=1,\ d_3=\frac12$,\ and
\begin{equation*}
\begin{aligned}
y^1=&e^{\pi i x_4}(e^{2\pi i x_1}+e^{2\pi i(x_2-x_1)}
+e^{2\pi i(x_3-x_2)}+e^{-2\pi ix_3});\\
y^2=&e^{2\pi i x_4}(e^{2\pi i x_2}+e^{-2\pi i x_2}
     +e^{2\pi i (x_1-x_3)}+e^{-2\pi i(x_1-x_3)}+\\
     &+e^{2\pi i (x_1+x_3-x_2)}+e^{-2\pi i(x_1+x_3-x_2)} );\\
y^3=&e^{\pi i x_4}(e^{2\pi i x_3}+e^{2\pi i(x_1-x_2)}
+e^{2\pi i(x_2-x_3)}+e^{-2\pi ix_1});\\
y^4=&2\pi i x_4.
\end{aligned}
\end{equation*}
The metric $\ (\ ,\ )^{\sptilde}$\ has the form
$$ 
((dx_i,dx_j)^{\sptilde})=\frac1{16\pi^2}\pmatrix
3&2&1&0\\ 2&4&2&0 \\
1&2&3&0\\ 
0&0&0&-4
\endpmatrix.
$$
The flat coordinates
$\
t_1=y^1,\ \  t_2=y^2,\ \ t_3=y^3,\ \ t_4=y^4
$,\
the intersection form is given by
$$
(g^{ij})=\pmatrix 2 t_2-\frac12 t_1^2
&3t_3e^{t_4}& 4e^{t_4}&\frac12t_1\\ 
3 t_3e^{t_4}&2t_1 t_3 e^{t_4}+4e^{2t_4}
&3t_1 e^{t_4}&t_2\\  4 e^{t_4}&3t_1 e^{t_4}&2t_2-\frac12 t_3^2
&\frac12 t_3\\ 
\frac12t_1&t_2&\frac12t_3&1
\endpmatrix.
$$
The free energy
$$
F=\frac14 t_1^2 t_2+\frac12 t_2^2 t_4+\frac1{4} t_2 t_3^2
-\frac1{96} t_1^4
-\frac1{96} t_3^4+t_1 t_3 e^{t_4}+\frac12 e^{2t_4},
$$
and the Euler vector field is given by
$$
E=\frac12 t_1\pal_1+t_2\pal_2+\frac12 t_3\pal_3+\pal_4.
$$
\vskip 0.5cm
\noindent {\bf {Example 2.7.}}\ \ Let $R$ be the root system $B_3$,
\ then $\ k=2,\ d_1=1,\ d_2=2,\ d_3=1$,\ and
\begin{equation*}
\begin{aligned}
&y^1=2 e^{2\pi i x_4}(\cos(x_1)+\cos(x_2-x_1)+\cos(2x_3-x_2));\\
&y^2=4 e^{4\pi i x_4}(\cos(x_1)\cos(x_2-x_1)+\cos(x_1)\cos(2x_3-x_2)+
          \cos(x_2-x_1)\cos(2x_3-x_2));\\
&y^3=8 e^{2\pi i x_4}(\cos(\frac{x_1}2)\cos(\frac{x_2-x_1}2)
\cos(\frac{2x_3-x_2}2));\\ 
&y^4=2\pi i x_4.
\end{aligned}   
\end{equation*}
The metric $\ (\ ,\ )^{\sptilde}$\ has the form
$$
((dx_i,dx_j)^{\sptilde})=\frac1{4\pi^2}\pmatrix
1&1&\frac12&0\\ 1&2&1&0\\
\frac12&1&\frac34&0\\
0&0&0&-\frac12
\endpmatrix.
$$
The flat coordinates
$\
t_1=y^1+2e^{y^4},\ \  
t_2=y^2+2 y_1 e^{y^4}+6 e^{2y^4},\ \ t_3=y^3,\ \ t_4=y^4 $,\
the elements of the intersection form are given by
\begin{equation*}
\begin{aligned}
&g^{11}=2 t_2-\frac12 t_1^2+4 e^{2t_4},\\
&g^{12}=3 t_3^2 e^{t_4}+6 t_1e^{2t_4},\\
&g^{13}=4 t_3 e^{t_4},\\
&g^{14}=\frac12 t_1,\\
&g^{22}=2t_1t_3^2 e^{t_4}+4 t_1^2 e^{2t_4}+8 t_3^2 e^{2t_4}+8 e^{4t_4},\\
&g^{23}=3 t_1t_3e^{t_4}+6 t_3e^{2t_4},\\
&g^{24}=t_2,\\
&g^{33}=t_2-\frac14 t_3^2+2t_1e^{t_4}+2e^{2t_4},\\
&g^{34}=\frac12 t_3,\\
&g^{44}=\frac12.
\end{aligned}
\end{equation*}
The free energy
$$
F=\frac14 t_2 t_1^2+\frac12 t_2 t_3^2+\frac12 t_2^2 t_4-\frac1{96}
t_1^4-\frac1{48} t_3^4+t_1 t_3^2 e^{t_4}+\frac12 t_1^2 e^{2t_4}+
t_3^2 e^{2t_4}+\frac14 e^{4t_4},
$$
and the Euler vector field is given by
$$
E=\frac12 t_1\pal_1+t_2\pal_2+\frac12 t_3\pal_3+\frac12\pal_4.
$$
\vskip 0.5cm
\noindent {\bf {Example 2.8.}}\ \ Let $R$ be the root system $C_3$,
\ then $\ k=3,\ d_1=1,\ d_2=2,\ d_3=3$,\ and
\begin{equation*}
\begin{aligned}
&y^1=e^{2\pi i x_4}(\xi_1+\xi_2+\xi_3);\\
&y^2=e^{4\pi i x_4}(\xi_1\xi_2+\xi_1\xi_3+\xi_2\xi_3);\\
&y^3=e^{6\pi i x_4}(\xi_1\xi_2\xi_3);\\
&y^4=2\pi i x_4,
\end{aligned}
\end{equation*}
where $\ \xi_1=e^{2\pi i x_1}+e^{-2\pi i x_1},\ \ 
\xi_2=e^{2\pi i(x_2-x_1)}+e^{-2\pi i(x_2-x_1)},\ \ 
\xi_3=e^{2\pi i (x_3-x_2)}+e^{-2\pi i (x_3-x_2)}$.\ 
The metric $\ (\ ,\ )^{\sptilde}$\ has the form 
$$ 
((dx_i,dx_j)^{\sptilde})=\frac1{4\pi^2}\pmatrix
1&1&1&0\\ 1&2&2&0\\
1&2&3&0\\ 
0&0&0&-\frac13
\endpmatrix.
$$
The flat coordinates
$\
t_1=y^1,\ \  t_2=y^2-\frac16 (y^1)^2+3 e^{2y^4},\ \ t_3=y^3+2 y^1 
e^{2y^4},\ \ t_4=y^4 $,\
the elements of the intersection form are given by
\begin{equation*}
\begin{aligned}
&g^{11}=-\frac13 t_1^2+2t_2+6e^{2t_4},\\
&g^{12}=3t_3+\frac1{18} t_1^3-t_1t_2+3t_1e^{2t_4},\\
&g^{13}=\frac43t_1^2 e^{2t_4}+8t_2 e^{2t_4},\\
&g^{14}=\frac13 t_1,\\
&g^{22}=-\frac1{54} t_1^4+\frac29 t_1^2t_2-\frac23t_2^2
+2t_1^2e^{2t_4}+4t_2e^{2t_4}+6e^{4t_4},\\
&g^{23}=\frac 59 t_1^3 e^{2t_4}+\frac{10}3t_1 t_2 e^{2t_4}
+10t_1e^{4t_4},\\
&g^{24}=\frac23 t_2,\\
&g^{33}=\frac19 t_1^4 e^{2t_4}+\frac 43 t_1^2 t_2 e^{2t_4}+4 t_2^2 e^{2t_4}
+8 t_1^2 e^{4t_4}+12 e^{6t_4},\\
&g^{34}=t_3,\\
&g^{44}=\frac13.
\end{aligned}
\end{equation*}
The free energy
\begin{equation*}
\begin{aligned}
F=&\frac13 t_1 t_2 t_3+\frac12 t_3^2 t_4+\frac1{18} t_2^3
-\frac1{36}t_1^2 t_2^2+\frac1{648}t_1^4 t_2-
\frac1{19440} t_1^6+\\
&\frac16 t_1^2t_2 e^{2t_4}+
\frac1{72} t_1^4 e^{2t_4}+\frac12 t_2^2 e^{2t_4}+
\frac14 t_1^2 e^{4t_4}+\frac16 e^{6t_4},
\end{aligned}
\end{equation*}
and the Euler vector field is given by
$$
E=\frac13 t_1\pal_1+\frac23 t_2\pal_2+t_3\pal_3+\frac13\pal_4.
$$
\par
\section{Groups $\ \tw^{(k)}(A_l)$\ and the spaces of 
trigonometric polynomials}
A trigonometric polynomial of one variable of bidegree $ (k,m)$ is 
a function of the form
\begin{equation*}
\begin{aligned}
\lm(\phi)=a_0e^{i k\phi}+&a_1 e^{i(k-1)\phi}+\cdots+a_k+a_{k+1} e^{-i\phi}
+\cdots+a_{k+m} e^{-im\phi},\\
& a_0,\dots,a_{k+m}\in \bold C,\quad a_0 a_{k+m}\ne 0.
\end{aligned}
\end{equation*}
We will usually normalize $\lm(\phi)$ by the condition
$\ a_0=1$.\par
We denote $\ M_{k,m}$\ the affine space of normalized trigonometric
polynomials. Equivalently, $ M_{k,m}$\ coincides with certain covering
of the space of rational functions with two poles of the orders $k$
and $m$ respectively. Geometry of these spaces was described in 
[D] as a part of general differential geometry
of Hurwitz spaces of branched coverings over $\bar {\bold C}$
(our space $M_{k,m}$ in the notations of [D] is $\hat {M}_{0;k-1,m-1}$).
Recall that, according to this paper, the space $M_{k,m}$ carries
a natural structure of Frobenius manifold. The invariant inner 
product of two vectors $\ \pal',\ \pal''$\ tangent to $M_{k,m}$
at a point $\ \lm(\phi)$\ equals 
\begin{equation}
<\pal',\pal''>_{\lm}=(-1)^{k+1}\sum_{|\lm |<\infty} \res_{d\lm=0}
\frac {\pal'(\lm(\phi)d\phi)\ \pal''(\lm(\phi)d\phi)}{d\lm(\phi)}.
\tag{3.1}
\end{equation}
In this formula the derivatives $\ \pal'(\lm(\phi) d\phi),\ 
\pal''(\lm(\phi) d\phi)$\ are to be calculated keeping $\ \phi$\ fixed.
The intersection form is given by the formula
\begin{equation}
(\pal',\pal'')_{\lm}=-\sum_{|\lm |<\infty}\res_{d\lm=0}  
\frac {\pal'(\log\lm(\phi)d\phi)\ \pal''(\log \lm(\phi)d\phi)}
{d\log\lm(\phi)}.
\tag{3.2}
\end{equation}
The discriminant $\ \cal \Sigma\subset M_{k,m}$\ consists of all functions
$\ \lm(\phi)$\ which fail to have all simple roots of the equation
$\ \lm(\phi)=0$.\ The intersection form is defined only outside
$\cal\Sigma$.\par
The formulae (3.1), (3.2) uniquelly determine multiplication of tangent
vectors on $M_{k,m}$ assuming that the Euler vector field $E$ has the
form
\begin{equation}
E=\sum_{j=1}^{k+m} \frac j k a_j\frac{\pal}{\pal a_j}.\tag{3.3}
\end{equation}
For any three tangent vectors $\ \pal',\ \pal'',\ \pal'''$  \
to $M_{k,m}$\ we obtain
\begin{equation}
<\pal' \cdot \pal'',\pal'''>_{\lm}=
-\sum_{|\lm |<\infty}\res_{d\lm=0}
\frac {\pal'(\lm(\phi)d\phi)\ \pal''(\lm(\phi)d\phi)\ 
\pal'''(\lm(\phi) d\phi)}{d\lm(\phi)\ d\phi}.\tag{3.4}
\end{equation}
The canonical coordinates $\ u_1,\dots,u_{k+m}$\ for this multiplication
are the critical values of $\ \lm(\phi)$:
\begin{equation}
\frac{\pal}{\pal u_\al} \cdot \frac{\pal}{\pal 
u_\beta}=\delta_{\al\beta}\frac{\pal}{\pal u_\al},\tag{3.5}
\end{equation}
(see [D] for details).\par
In this section we will show that the space $M_{k,m}$ as a Frobenius
manifold is isomorphic to the orbit space of our extended affine
Weyl group $\ \tw^{(k)}(A_l)$\ for $\ l=k+m-1$.\par
We start with factorizing the trigonometric polynomial
\begin{equation}
\lm(\phi)=e^{-im\phi}\prod_{\al=1}^{k+m}(e^{i\phi}-e^{i\phi_\al}).\tag{3.6}
\end{equation}
\begin{Lem} The map 
\begin{equation}
(x_1,\dots,x_{k+m})\mapsto (\phi_1,\dots,\phi_{k+m}),\tag{3.7}
\end{equation}
where
\begin{equation*}
\begin{aligned}
&\phi_1=2\pi (x_1+\frac{m}{k+m} x_{k+m}),\quad
\phi_j=2\pi (x_j-x_{j-1}+\frac{m}{k+m} x_{k+m}),\\
&\phi_{k+m}=2\pi (-x_{k+m-1}+\frac{m}{k+m} x_{k+m}),\quad
j=2,\dots,k+m-1\end{aligned}\tag{3.8}
\end{equation*}
establishes a diffeomorphism of the orbit space of the group 
$\ \tw^{(k)}(A_{k+m-1})$\ to the space of normalized trigonometric 
polynomials.\end{Lem}
\begin{pf} From the explicit formulae (1.6), (1.10), (2.1) and
(3.8) it follows that in the coordinates $\ y^1,\dots,y^{k+m}$\
the map (3.7) has the form
\begin{equation*}
\begin{aligned}
&a_1=-y^1,\\
\ &\cdots\\
&a_k=(-1)^k y^k,\\
&a_{k+1}=(-1)^{k+1} y^{k+1} \exp(y^{k+m}),\\
\ &\cdots\\
&a_{k+m-1}=(-1)^{k+m-1} y^{k+m-1} \exp((m-1) y^{k+m}),\\
&a_{k+m}=(-1)^{k+m} \exp(m y^{k+m}).
\end{aligned}\tag{3.9}
\end{equation*}
Lemma is proved.\end{pf}\par
According to this Lemma, our group $\ \tw^{(k)}(A_l)$\ describes monodromy
of logarithms of the roots of a trigonometric polynomial along closed
loops in the space of coefficients nonintersecting the discriminant
$\cal\Sigma$.\par
\begin{Thm} The diffeomorphism (3.7) is an isomorphism 
of Frobenius manifolds.\end{Thm}
\begin{pf} Since the Euler vector fields (2.26) and  (3.3) coincide,
it suffices to prove that the intersection form (3.2) coincides with
the intersection form of the orbit spaces,
and the metric (3.1) coincides with the metric (2.9).\par
Let's denote the roots of $\ \lm'(\phi)$\ by $\ \psi_j,\ 1\le j\le k+m$.\ 
Then we have
\begin{equation}
\lm'(\phi)=k i e^{-im\phi}\prod_{\al=1}^{k+m} (e^{i\phi}-e^{i\psi_\al}).
\tag{3.10}
\end{equation}
We define $\ u_\al=\lm(\psi_\al),\ 1\le \al\le k+m$,\ then
\begin{equation}
\pal_{u_\al}\lm(\phi)|_{\phi=\psi_\beta}=\delta_{\al\beta}.\tag{3.11}
\end{equation}
By using (3.10), (3.11) and the 
Lagrange interpolation formula we obtain
\begin{equation}
\pal_{u_\al}\lm(\phi)=\frac{i e^{i\psi_\al}\lm'(\phi)}
{(e^{i\phi}-e^{i\psi_\al})
\lm''(\psi_\al)}.\tag{3.12}
\end{equation}
Formulae (3.6) and (3.12) lead to
\begin{equation}
-\sum_{a=1}^{k+m}i\pal_{u_\al}\phi_a e^{i\phi_a}\frac{\lm(\phi)}
{e^{i\phi}-e^{i\phi_a}}=\frac{i 
e^{i\psi_\al}\lm'(\phi)}{(e^{i\phi}-e^{i\psi_\al})
\lm''(\psi_\al)}.\tag{3.13}
\end{equation}
By putting $\ \phi=\phi_\beta$\ in the above formula we obtain
\begin{equation}
\pal_{u_\al}\phi_{\beta}=-\frac{i e^{i\psi_{\al}}}{(e^{i\phi_{\beta}}-
e^{i\psi_{\al}})\lm''(\psi_\al)}.\tag{3.14}
\end{equation}
We denote
\begin{equation}
\ta_1=x_1,\ \ \ta_j=x_j-x_{j-1},\ j=2,\dots,k+m-1,\ \ 
\ta_{k+m}=x_{k+m}.\tag{3.15}
\end{equation}
It follows from (3.8) and (3.14) that
\begin{equation*}
\begin{aligned}
&\frac{\pal \ta_\beta}{\pal u_\al}=\frac{1}{2\pi i} 
\frac{e^{i\psi_\al}}{(e^{i\phi_{\beta}}-
e^{i\psi_{\al}})\lm''(\psi_\al)}-\frac{m}{k+m}\frac{\pal\ta_{k+m}}{\pal u_\al},
\quad 1\le\beta\le k+m-1,\\
&\frac{\pal\ta_{k+m}}{\pal u_\al}=\frac1{2m\pi i}\sum_{a=1}^{k+m}
\frac{e^{i\psi_\al}}{(e^{i\phi_{a}}-
e^{i\psi_{\al}})\lm''(\psi_\al)}.
\end{aligned}\tag{3.16}
\end{equation*}
From (3.2) and (3.11) we obtain
\begin{equation}
\tilde g_{\al\beta} 
:=(\pal_{u_\al},\pal_{u_\beta})_{\lm}=-\frac{\delta_{\al\beta}}
{u_{\al}\lm''(\psi_\al)}.\tag{3.17}
\end{equation}
In a similar way we can compute the inner product of the vectors 
$\ \pal_{u_{\al}}$\ w.r.t. the bilinear form (3.1), the result reads
\begin{equation}
\tilde\eta_{\al\beta} :=
<\pal_{u_\al},\pal_{u_\beta}>_{\lm}=(-1)^{k+1}\frac{\delta_{\al\beta}}
{\lm''(\psi_\al)}.\tag{3.18}
\end{equation}
We observe now that the vector field $\ e=\frac{\pal}{\pal y^k}$ \
in the coordinates $\ a_1,\dots, a_{k+m}$\ coincides with
\begin{equation*}
 e=(-1)^k \frac{\pal}{\pal a^k}.
\end{equation*}
This follows from (3.9). Shift
$$
a_k\mapsto a_k+c
$$
produces the correspondent shift
$$
u_i\mapsto u_i+c,\quad i=1,\dots,k+m
$$
of the critical values. This shift does not change the critical points
$\ \psi_\al$\ neither the values of the second derivative 
$\ \lm''(\psi_\al)$. \ So 
\begin{equation}
\cal L_e \tilde g^{\al\beta}=\cal L_e (-u_\al 
\lm''(\psi_\al)\delta_{\al\beta})
=(-1)^{k+1}\lm''(\psi_\al)\delta_{\al\beta}=\tilde\eta^{\al\beta},\tag{3.19}
\end{equation}
where $\ (\tilde g^{\al\beta})=(\tilde g_{\al\beta})^{-1},\ \ \
(\tilde \eta^{\al\beta})=(\tilde \eta_{\al\beta})^{-1}$.\par
Now we proceed to the computation of the bilinear form (3.2) in the 
coordinates $\ x_1,\dots, x_{k+m}$\ of (3.7) (or, equivalently in the 
coordinates $\ \theta_1,\dots,\theta_{k+m}$ \ of the form (3.15)).
It turns out that this coincides with the form $(\ ,\ )^{\sptilde}$ defined
in Section 2 above.\par 
We will use the following identity:
\begin{equation}
\sum_{\al=1}^{k+m}\frac{u_\al e^{2i\psi_\al}}{(e^{i\phi_a}-e^{i\psi_\al})
(e^{i\phi_b}-e^{i\psi_\al})\lm''(\psi_\al)}=\delta_{ab}-\frac1k.\tag{3.20}
\end{equation}
In fact the left-hand side of (3.20) equals
\begin{equation*}
\begin{aligned}
&\sum_{\al=1}^{k+m}\res_{\phi=\psi_\al}\frac{\lm(\phi) 
e^{2i\phi}}{(e^{i\phi}-e^{i\phi_a})
(e^{i\phi}-e^{i\phi_b})\lm'(\phi)}\\
=&\sum_{\al=1}^{k+m}\res_{v=e^{i\psi_\al}}\frac{\lm(-i \log v)\ v^2   
\frac1{i v}}{(v-e^{i\phi_a})         
(v-e^{i\phi_b})\lm'(-i\log v)}\\
=&(\res_{v=e^{i\phi_a}}+\res_{v=e^{i\phi_b}}+\res_{v=\infty})
\frac{i\lm(-i \log v)\ v}{(v-e^{i\phi_a})          
(v-e^{i\phi_b})\lm'(-i\log v)}\\    
=&\delta_{ab}-\frac1k.
\end{aligned}
\end{equation*}
By using (3.16), (3.17) and (3.20) we obtain
\begin{equation*}
\begin{align}
&(d\ta_{k+m},d\ta_{k+m})_{\lm}=\sum_{\al=1}^{k+m}\frac{1}{\tilde 
g_{\al\al}(u)}\frac{\pal \ta_{k+m}}{\pal u_\al}
\frac{\pal \ta_{k+m}}{\pal u_\al} \notag \\
=&\sum_{\al=1}^{k+m}(-u_\al \lm''(\psi_\al))\frac1{(2m\pi i)^2}
\sum_{a,b=1}^{k+m}\frac{e^{2i\psi_\al}}{(e^{i\phi_a}-e^{i\psi_\al})
(e^{i\phi_b}-e^{i\psi_\al})\lm''(\psi_\al)^2}\notag
\\
=&\frac1{4m^2\pi^2}\sum_{\al=1}^{k+m}\sum_{a,b=1}^{k+m}
\frac{u_\al e^{2i\psi_\al}}{(e^{i\phi_a}-e^{i\psi_\al})
(e^{i\phi_b}-e^{i\psi_\al})\lm''(\psi_\al)}\notag
\\
=&\frac1{4m^2\pi^2}\sum_{a,b=1}^{k+m}(\delta_{ab}-\frac1k)\notag \\
=&\frac1{4m^2\pi^2}(k+m-\frac{(k+m)^2}{k})\notag \\
=&-\frac1{4\pi^2}\frac{k+m}{mk}=-\frac1{4\pi^2 d_k}.\tag{3.21}
\end{align}
\end{equation*}
For any $\ 1\le \al\le k+m-1$,\ it follows from (3.16),
(3.17), (3.20) and (3.21)  that
\begin{equation*}
\begin{align}
&(d\ta_{k+m},d\ta_{\al})_{\lm}=\sum_{a=1}^{k+m}\frac{1}{\tilde
g_{aa}(u)}\frac{\pal \ta_{k+m}}{\pal u_a}
\frac{\pal \ta_\al}{\pal u_a}\notag \\
=&\sum_{a=1}^{k+m}\frac{1}{\tilde  
g_{aa}(u)}\frac{\pal \ta_{k+m}}{\pal u_a}
(\frac{1}{2\pi i} \frac{e^{i\psi_a}}{(e^{i\phi_{\al}}-
e^{i\psi_{a}})\lm''(\psi_a)}-\frac{m}{k+m}\frac{\pal\ta_{k+m}}{\pal u_a})
\notag \\
=&-\frac{m}{k+m} (-\frac1{4\pi^2}\frac{k+m}{mk})+\frac1{2\pi i}
\sum_{a=1}^{k+m}\frac{1}{\tilde
g_{aa}(u)}\frac{\pal \ta_{k+m}}{\pal 
u_a}\frac{e^{i\psi_a}}{(e^{i\phi_{\al}}-    
e^{i\psi_{a}})\lm''(\psi_a)}\notag \\
=&\frac1{4k\pi^2}+\frac1{4m\pi^2}\sum_{a,b=1}^{k+m}
\frac{u_a e^{2i\psi_a}}{(e^{i\phi_{\al}}-
e^{i\psi_{a}})(e^{i\phi_b}-
e^{i\psi_{a}})\lm''(\psi_a)}\notag \\
=&\frac1{4k\pi^2}+\frac1{4m\pi^2}\sum_{b=1}^{k+m}(\delta_{\al b}-\frac1k)
\notag \\
=&\frac1{4k\pi^2}+\frac1{4m\pi^2}(1-\frac{k+m}k)=0.\tag{3.22}
\end{align}
\end{equation*}
Finally, for any $\ 1\le \al, \beta\le k+m-1$,\ by using (3.16), 
(3.17) and (3.20)--(3.22) we obtain
\begin{equation*}
\begin{align}
&(d\ta_\al,d\ta_{\beta})_{\lm}=\sum_{a=1}^{k+m}\frac{1}{\tilde
g_{aa}(u)}\frac{\pal \ta_\al}{\pal u_a}
\frac{\pal \ta_\beta}{\pal u_a}\notag \\
=&\sum_{a=1}^{k+m}\frac{1}{\tilde
g_{aa}(u)}
(\frac{1}{2\pi i} \frac{e^{i\psi_a}}{(e^{i\phi_{\al}}-
e^{i\psi_{a}})\lm''(\psi_a)}-\frac{m}{k+m}
\frac{\pal\ta_{k+m}}{\pal u_a})\times\notag \\
&\ \times (\frac{1}{2\pi i} \frac{e^{i\psi_a}}{(e^{i\phi_{\beta}}-    
e^{i\psi_{a}})\lm''(\psi_a)}-\frac{m}{k+m}\frac{\pal\ta_{k+m}}{\pal u_a})
\notag \\
=&\sum_{a=1}^{k+m}\frac{1}{\tilde
g_{aa}(u)}\frac{1}{(2\pi i)^2}
\frac{e^{2i\psi_a}}{(e^{i\phi_\al}-e^{i\psi_a})
(e^{i\phi_\beta}-e^{i\psi_a})\lm''(\psi_\al)^2}-\notag \\
&\ -\frac{m}{2\pi i(k+m)}\sum_{a=1}^{k+m}\frac{1}{\tilde
g_{aa}(u)}
\frac{e^{i\psi_a}}{(e^{i\phi_{\al}}-
e^{i\psi_{a}})\lm''(\psi_a)}
\frac{\pal\ta_{k+m}}{\pal u_a}\notag \\
=&\frac1{4\pi^2}(\delta_{\al\beta}-\frac1k)-\frac{m}{2\pi i(k+m)}
\frac{2\pi i}{4m\pi^2}(1-\frac{k+m}k)\notag \\
=&\frac1{4\pi^2}(\delta_{\al\beta}-\frac1{k+m}).\tag{3.23}
\end{align}
\end{equation*}
Now the coincidence of 
$\ (dx_i,dx_j)_{\lm}$\ and
$\ (dx_i,dx_j)^{\sptilde}$\ follows easily from (3.21)--(3.23). Hence the
intersection form (3.2) coincides with the intersection form of the 
orbit spaces. The coincidence of the metric (3.1) with the 
metric (2.9) follows (3.19). 
Theorem is proved.\end{pf}
\par
We construct now the flat coordinates $\ t^1,\dots,t^{k+m}$\ on
the space of trigonometric polynomials (essentially, following [D]).
Let's define
\begin{equation*}
\begin{aligned}
&t^\mu=(-1)^{\mu+1}\frac{k i}{\mu}\res_{e^{i\phi}=\infty} \lm(\phi)^{\frac 
\mu k}\ d\phi, \quad 1\le \mu\le k-1\\
&t^{k+m-\mu}=(-1)^{\mu}\frac{m i}{\mu}\res_{e^{-i\phi}=\infty}
[(-1)^{k+m}\lm(\phi)]^{\frac {\mu}{m}}\ d\phi,
\quad 1\le \mu\le m\\
&t^{k+m}=y^{k+m}.
\end{aligned}\tag{3.24}
\end{equation*}
From the above definition we have
\begin{equation*}
\begin{aligned}  
&t^\mu=y^{\mu}+f_\mu (y^1,\dots,y^{\mu-1}),\quad 1\le \mu\le k-1\\
&t^{k+m-\mu}= y^{k+m-\mu}+h_{\mu}(y^{k+m-1},\dots,y^{k+m-\mu+1}),
\quad 1\le i\le m-1,\\
&t^{k}= y^k,\quad t^{k+m}=y^{k+m},
\end{aligned}\tag{3.25}
\end{equation*}
where $f$'s and $h$'s are some polynomials, and the relation
between $y$'s and $a$'s is given in (3.9).\par
\begin{Prop}
The variables $t^\mu$\  are the flat coordinates for the metric (3.1).
\end{Prop}
\begin{pf}
Let's denote
$$
\xi=\lm(\phi)^{\frac 1 k},\quad \eta=[(-1)^{k+m}\lm(\phi)]^{\frac 1{m}},
$$
it follows from (3.24) that when $\ e^{i\phi}$\ tends to infinity we 
have 
$$
\phi=-i\log\xi-\frac ik[\frac {t^1}{\xi}-\frac{t^2}{\xi^2}+
\cdots+(-1)^{k}\frac{t^{k-1}
}{\xi^{k-1}}]+\cal{O}(\frac 1{\xi^k}),
$$  
and when $\ e^{-i\phi}$\ tends to infinity we have
$$
\phi=i\log\eta-i\ t^{k+m}+\frac{i}{m} [\frac {t^{k+m-1}}{\eta}
-\frac{t^{k+m-2}} 
{\eta^2}+\cdots+(-1)^{m-1}\frac{t^{k}
}{\eta^{m}}]+\cal{O}(\frac 1{\eta^{m}}),
$$
By using the ``thermodynamical identity'' [D] 
$$
\pal_{t^\mu}(\lm d\phi)_{\phi=\text {constant}}
=-\pal_{t^\mu}(\phi d\lm)_{\lm=\text {constant}}
$$
we obtain
\begin{equation*}
\begin{aligned}
&\pal_{t^\al}(\lm(\phi)d\phi)=
\cases (-1)^{\al+1}\ i\xi^{k-\al-1}d\xi+\cal O(\frac1{\xi})d\xi,& 
e^{i\phi}\to \infty\\
       \cal O(\frac 1{\eta}) d\eta,& e^{-i\phi}\to \infty\endcases\\ 
&\pal_{t^{k+m-\beta}}(\lm(\phi)d\phi)=
\cases \cal O(\frac 1{\xi})d\xi,& e^{i\phi}\to \infty\\
(-1)^{k+m-\beta}\ i\eta^{m-\beta-1}\ d\eta+\cal O(\frac 1{\eta})d\eta,& 
e^{-i\phi}\to \infty \endcases \\ 
&\pal_{t^{k+m}}(\lm(\phi)d\phi)=
\cases \cal O(\frac1{\xi})d\xi,& e^{i\phi}\to \infty\\ 
     (-1)^{k+m} m i \eta^{m-1}d\eta+ \cal O(\frac 1{\eta}) d\eta,&
      e^{-i\phi}\to \infty\endcases
\end{aligned}
\end{equation*}
where $1\le \al\le k-1$,\ $1\le \beta\le m$. Thus for
 $\ 1\le \al,\ \beta\le k-1$\ we have
\begin{equation*}
\begin{aligned}
&<\pal_{t^{\al}},\pal_{t^\beta}>_{\lm}=(-1)^k (\res_{e^p=\infty}
+\res_{e^{-p}=\infty})
\frac{\pal_{t^\al}(\lm(\phi) d\phi)\ \pal_{t^\beta}(\lm(\phi)
d\phi)}{d\lm(\phi)}\\ 
&=(-1)^k (-1)^{\al+\beta+1}\res_{e^{p}=\infty}\frac{\xi^{k-\al-1}d\xi \ 
\xi^{k-\beta-1}
d\xi}{d\xi^k}=\frac{1} k \delta_{\al+\beta,k};
\end{aligned}
\end{equation*}
similarily, we have
\begin{equation*}
\begin{aligned}
&<\pal_{t^\al},\pal_{t^{k+m-\beta}}>_{\lm}=0,\quad 1\le \al\le k-1,\
1\le \beta\le m,\\
&<\pal_{t^{k+m-\al}},\pal_{t^{k+m-\beta}}>_{\lm}=
\frac{1}m \delta_{\al+\beta,m},\quad 1\le \al, \beta\le m,\\
&<\pal_{t^{k+m}},\pal_{t^{\al}}>_{\lm}= \delta_{\al,k},
\quad 1\le \al\le k+m.
\end{aligned}
\end{equation*}
Proposition is proved.\end{pf}
\par
We conclude this section with an example of topological application
of Theorem 3.1.\  Let $M_{k,m}^0$ be the subspace consisting of all
trigonometric polynomials having $k+m$ pairwise distinct critical 
values $\ u_1,\dots,u_{k+m}$.
What is the topology of $M_{k,m}^0$ ?
\newline
Particularly, how to compute 
the number $N(k,m)$ of trigonometric polynomials of the bidegree 
$(k,m)$ with given pairwise distinct nondegenerate critical values?
\par
More generally, for a $n$-dimensional complex Frobenius manifold $M$
satisfying semi-simplicity condition (i.e., the algebra on $\ \text {T}_t M$
\ is semisimple for a generic $t$) we may consider the open subset 
$\ M^0$\ consisting of all points $\ t$\ such that the eigenvalues 
$\ u_1(t),\dots,$ $u_n(t)$\ of the operator of multiplication by the Euler 
vector field are pairwise distinct. According to [D], on $\, M^0$\, the
eigenvalues can serve as local coordinates. They are called 
{\it canonical coordinates} since the multiplication table of tangent 
vectors takes the very simple form (3.5) in the coordinates 
$\,(u_1,\dots,u_n)$.\, [As we have explained above, for the space 
$\, M_{k,m}$\, of trigonometric polynomials $\, \lm(\phi)$\, the
canonical coordinates coincide with the critical values of 
$\, \lm(\phi)$].\, According to [D], the map
\begin{equation*}
\begin{aligned}
&M^0\to (\bold C^n\setminus {\text{diagonals}})/S_n\\
&t \mapsto (u_1(t),\dots,u_n(t))\,{\text{modulo permutations}}
\end{aligned}
\end{equation*}
establishes an equivalence between $\,M^0$\, and the space of 
isomonodromy deformations of certain linear differential operator
with rational coefficients. What is the topology of this space?
Particularly, how to compute the number $N(M)$ of the points
$\,t\in M^0$\, with given pairwise distinct canonical coordinates
$\, u_1,\dots,u_n$\,?\par 
To find this number we will compute the degree of the map
$$
M^0\to \bold C^n
$$
given by the formula
\begin{equation}
t\mapsto (b_1(t),\dots,b_n(t)),\tag{3.26}
\end{equation}
where $\ b_1(t),\dots,b_n(t)$\ are coefficients of the characteristic 
polynomial
\begin{equation*}
\begin{aligned}
&(-1)^n\,[u^n+b_1(t)\,u^{n-1}+\cdots+b_n(t)] :=\\
&\det((E(t)\cdot )-u)=(\det<\ ,\ >_t)^{-1}\det((\ ,\ )_t-u<\ ,\ 
>_t)=\prod_{i=1}^n (u_i(t)-u),
\end{aligned}
\end{equation*}
here $\ (\ ,\ )_t$\ and $\ <\ ,\ >_t$\ are the intersection form and
the invariant metric respectively considered as bilinear forms on
$\ \text{T}_t^* M$.\ We call it generalized Looijenga-Lyashko map 
(LL-map) of the Frobenius manifold (cf. [A3, Lo, Ly]).\par
The degree of LL-map can be computed easily in the case of polynomial
Frobenius manifolds. More precisely, let the free energy $F$ defining
the Frobenius structure in the flat coordinates $\ t^1,\dots,t^{n}$\ has
the form
$$
F=\text {cubic term}+G(t^1,\dots,t^p,q^1,\dots,q^r),
$$
where $\ G$\ is a polynomial,
$$
p+r=n,\quad q^i=\exp t^{p+i},\quad i=1,\dots,r,
$$
and the degrees of the variables $\ t^{p+1},\dots, t^n$\ are equal to
zero. Let us assume that all the degrees
$$
\deg t^i,\quad 1\le i\le p,\quad \deg q^i,\quad 1\le i\le r
$$
are positive. Then LL-map is a polynomial map $\ M\to \bold C^n$.\
We recall  that the degrees are normalized in such a way that
$\ \deg t^1=1$.\ Then the degrees of the canonical coordinates
$\ u_1,\dots,u_n$\ are equal to $1$ [D]. Hence the weighted degrees of 
the functions $\ b_1(t),\dots,b_n(t)$\ are equal to $\ 1,\dots,n$\
respectively.\par
Thus to compute the degree of the LL-map (3.26) we can use the graded
Bezout theorem. We obtain
\begin{equation}
N(M)=
\deg {\text {LL}}=\frac{n!}{\deg t^1\cdots\deg t^p\deg q^1\cdots\deg q^r}.
\tag{3.27}
\end{equation}\par
In the case of the orbit space of the group $\ \tw^{(k)}(A_{k+m-1})$\
the degrees of the variables $\ t^1,\dots,t^{k+m-1}$\ are expressed
via inner products of fundamental weights (i.e., via entries of inverse
of the  $\ A_{k+m-1}$\ Cartan  matrix):
$$
\deg t^j=\frac{(\w_j,\w_k)}{(\w_k,\w_k)},\quad j=1,\dots,k+m-1,
$$
and 
$$
\deg\exp t^{k+m}=\frac 1{(\w_k,\w_k)}.
$$
We arrive at the following formula for the degree of LL-map of the
Frobenius manifold $M_{k,m}$:
$$
\deg\text{LL}=\frac{(k+m)!\ 
(\w_k,\w_k)^{k+m}}{\prod_{j=1}^{k+m-1}(\w_j,\w_k)}.
$$
Using the explicit expression for $(\w_j,\w_k)$ (see Table 2 above)
we derive the formula for the number $\ N(k,m)$\,
(obtained first by Arnold in
[A2])
$$
N(k,m)=k^k m^m\frac{(k+m-1)!}{(k-1)!\ (m-1)!}.
$$
\par
We hope that our extended affine Weyl groups could be useful for other
problems arising in topological study of spaces of rational functions.
\par
\vskip 0.8cm
{\bf{Acknowledgments.}}\ \ One of the authors (B. D.) thanks 
V.I. Arnold and S.M. Natanzon for fruitful discussions, the author 
(Zhang) thanks K. Saito and P. Slodowy for valuable 
discussions. The 
authors thank the referee of the paper for the reference to [A3].\par
\newpage

{\bf Note added in proof.} In January of '97 after the article has been
submitted to the journal, the authors received an interesting paper of
P.Slodowy ``A remark on a recent paper by B.Dubrovin and Y.Zhang". In this
paper it is shown that our analogue of Chevalley theorem for extended
affine Weyl groups can be derived from the results of K.Wirthm\"uller
``Torus embeddings and deformations of simple space curves", {\sl Acta
Mathematica} {\bf 157} (1986) 159-241. This raises a natural question
(already formulated by P.Slodowy) to extend (if possible) our construction
of Frobenius structures
to the more
general setting of Wirthm\"uller. We hope to address the problem in
subsequent publications.

\end{document}